\documentclass[preprint]{aastex}
\usepackage{gensymb} 
\usepackage{graphicx}
\usepackage{natbib}

\shorttitle{DISCS: I. Taurus Protoplanetary Disk Data}
\shortauthors{\"Oberg et al.}

\begin{document}

\title{Disk Imaging Survey of Chemistry with SMA (DISCS): \\I. 
Taurus Protoplanetary Disk Data}

%\author{ }
%\affil{ }

\author{
Karin I. \"Oberg\altaffilmark{1}, 
Chunhua Qi}
\affil{Harvard-Smithsonian Center for Astrophysics, 60 Garden Street, Cambridge, MA 02138, USA} 

\and

\author{
Jeffrey K.J. Fogel, 
Edwin A. Bergin}
\affil{Department of Astronomy, University of Michigan, Ann Arbor, MI 48109, USA}

\and

\author{
Sean M. Andrews\altaffilmark{1}, 
Catherine Espaillat\altaffilmark{2}, 
Tim A. van Kempen, 
David J. Wilner}
\affil{Harvard-Smithsonian Center for Astrophysics, 60 Garden Street, Cambridge, MA 02138, USA}

\and

\author{Ilaria Pascucci}
\affil{Department of Physics and Astronomy, Johns Hopkins University, 3400 N. Charles Street, Baltimore, MD 21218, USA}

\altaffiltext{1}{Hubble Fellows}
\altaffiltext{2}{NSF Astronomy \& Astrophysics
Postdoctoral Fellow}

\begin{abstract}
Chemistry plays an important role in the structure and evolution of  protoplanetary disks, with implications for the composition of comets and planets. This is the first of a series of papers based on data from DISCS, a  Submillimeter Array survey of the chemical composition of protoplanetary disks.  The six Taurus sources in the program (DM Tau, AA Tau, LkCa 15, GM Aur, CQ Tau  and MWC 480) range in stellar spectral type from M1 to A4 and offer an  opportunity to test the effects of stellar luminosity on the disk  chemistry. The disks were observed in 10 different lines at $\sim$3\arcsec resolution and an rms of $\sim100$ mJy~beam$^{-1}$ at $\sim0.5$ km~s$^{-1}$. The four brightest lines are CO 2--1, HCO$^+$ 3--2, CN 2$_{3 \: 3/4/2}-1_{2 \: 2/3/1}$ and HCN 3--2 and these are  detected toward all sources (except for HCN toward CQ Tau). The weaker lines of CN 2$_{2 \: 2}-1_{1 \: 1}$,  DCO$^+$ 3--2, N$_2$H$^+$ 3--2, H$_2$CO 3$_{0 \: 3}-2_{0 \: 2}$ and 4$_{1 \: 4}-3_{1 \: 3}$ are detected toward two to three disks each, and DCN 3--2 only toward LkCa 15. CH$_3$OH $4_{2\: 1}-3_{1\: 2}$ and $c$-C$_3$H$_2$ are not detected. There is no obvious difference between the T Tauri and Herbig Ae sources  with regard to CN and HCN intensities. In contrast, DCO$^+$, DCN, N$_2$H$^+$ and H$_2$CO are detected only toward the  T Tauri stars, suggesting that the disks around Herbig Ae stars lack cold regions for long enough timescales to allow for efficient deuterium chemistry, CO freeze-out, and grain chemistry.
\end{abstract}


\section{Introduction}

Protoplanetary disks link protostars and (extra-)solar planetary systems 
physically and chemically. Understanding the chemical composition and 
evolution of disks thus provides constraints on the nature of molecules 
incorporated into planetesimals and planets. 
A variety of simple species, including the organic molecules CN, HCN, and
H$_2$CO, have been detected towards a handful of disks in unresolved studies 
\citep{Dutrey97, Thi04,Kastner08}, suggestive of an active chemistry. However, apart 
from CO and to some extent HCO$^+$, the chemistry is poorly constrained \citep{Pietu07}. 

Observations of the earlier stages of star formation and 
of comets, the possible remnants of our own protoplanetary disk, reveal a 
chemistry rich in simple and complex organics up to HCOC$_2$H$_5$ and 
(CH$_2$OH)$_2$ in size \citep[e.g.][]{vanDishoeck95, Crovisier04, Belloche09}. 
Pre-biotic pathways to chemical complexity thus exist. If theses pathways are active 
in disks or if the observed cometary complexity is instead a fossil remnant from earlier stages remains to be shown. Recent experiments suggest that the combination 
of icy grain mantles and UV irradiation efficiently produces the complex 
molecules found around protostars \citep{Oberg09d}, and surely appropriate 
conditions are common in disks as well \citep{vanZadelhoff03,Hersant09}.

The chemistry in disks is significantly more difficult to probe than in 
protostellar cores because of their order-of-magnitude smaller angular size, which
necessitates the use of (sub-)millimeter arrays to resolve the chemistry of 
the bulk of the disk material. The first observations of disk chemistry by \citet{Dutrey97} also revealed lower gas-phase abundances of most molecules compared to protostars. From a slightly larger sample consisting of four sources \citet{Thi04} showed that protoplanetary disks seem to generally contain orders of magnitude lower gas abundances compared to protostellar cores. This is reproduced by models of disks with a combination of 
freeze-out onto grains toward the disk midplane and photodissociation in the 
disk atmosphere \citep[e.g.][]{Aikawa99}. Gas phase molecules are only expected to be abundant in an intermediate zone that is warmer than common ice sublimation temperatures, but still deep 
enough into the disk to be partly protected from stellar and interstellar UV 
irradiation \citep[e.g.][]{Aikawa99, Bergin03}. The low molecular abundances result in
weak emission that requires long integration times at all existing 
(sub-)millimeter facilities. The investment required for interferometers 
in large part explains the small number of resolved chemistry studies of any species 
more complicated than HCO$^+$ \citep{Qi03, Dutrey04, Dutrey07, Qi08, Henning10}.

Despite these impediments, single dish studies of the T Tauri stars 
DM Tau, GG Tau, LkCa 15, TW Hya, V4046 Sgr and the Herbig Ae stars MWC 480 
and HD 163296 have provided some constraints on the chemistry of protoplanetary
disks \citep{Dutrey97,Thi04,Kastner08}. 
The species CN, HCN, DCN, HNC, CS, C$_2$H, H$_2$CO, HCO$^+$ and DCO$^+$ have 
been detected toward at least one of these objects, with CN/HCN ratios that 
only can be explained by high UV or X-ray fluxes penetrating into the disk -- CN is a photodissociation product of HCN. 
The variations in molecular abundances among different systems are significant; H$_2$CO is only 
detected toward DM Tau and LkCa 15, DCO$^+$ toward TW Hya, and HCN toward 
all T Tauri stars but not toward either of the Herbig Ae stars. Suggested reasons 
for these variations include higher photodissociation rates and a lack of cold chemistry products toward the more luminous Herbig Ae stars, different stages of 
grain growth in different disks, and different disk structures. Differences in 
the chemistry preceding the disk stage may also play a role. 

Resolved studies of disk chemistry are few but intriguing. Using the 
IRAM Plateau de Bure Interferometer, \citet{Dutrey07} reported low 
signal-to-noise N$_2$H$^+$ detections toward DM Tau and LkCa 15 and an 
upper limit toward MWC 480. Within the same project \citet{Schreyer08} tentatively detected 
HCN toward AB Aur. More recently \citet{Henning10} resolved C$_2$H emission toward DM Tau and LkCa 15, while no C$_2$H was detected toward the more luminous MWC 480. The difference between the T Tauri stars and the Herbig Ae star was explained by a combination of high UV and low X-ray fluxes toward MWC 480. Using the SMA, \citet{Qi08} spatially resolved the 
HCO$^+$ and DCO$^+$ emission toward TW Hydrae, revealing different 
distributions of these 
species -- DCO$^+$ is relatively more abundant at increasing radii out to 
90 AU, consistent with its origins from cold disk chemistry. HCN and DCN were 
both detected as well, but provide less constraints on the chemistry because 
of the weak DCN signal. Overall, the combination of small samples of diverse 
sources and even fewer resolved studies have so far prevented any strong 
constraints on the main source of chemical diversity in protoplanetary disks.

With the DISCS (Disk Imaging Survey of Chemistry with SMA) Submillimeter Array legacy program, we aim to produce a resolved, systematic survey
of chemistry toward protoplanetary disks spanning a range of spectral types and
disk parameters. The targeted molecules are the simple species that
previous studies suggested may be detectable (CO, HCO$^+$, DCO$^+$, CN, HCN, DCN,
N$_2$H$^+$, C$_3$H$_2$, H$_2$CO and CH$_3$OH). The initial survey contains six 
well known disks in the Taurus molecular clouds (DM Tau, AA Tau, LkCa 15, 
GM Aur, CQ Tau and MWC 480) with central stars that span spectral types M1 to 
A4. 
The disk sample is presented in $\S$\ref{sec:sample} with special attention to 
the properties that may affect the disk chemistry such as stellar luminosity, accretion rates,
disk size, disk structure and dust settling. 
The spectral set-ups and observational details are described in 
$\S$\ref{sec:obs}.
The channel maps toward one of the richest sources, moment maps for all disks 
in the most abundant species, and spectra of all detected lines toward all 
sources are shown in $\S$\ref{sec:res}. 
The detection rates as well as source-to-source variations are discussed in 
section $\S$\ref{sec:disc} followed by qualitative discussions on the origins 
of the observed chemical variations.

\section{The Disk Sample\label{sec:sample}}

\subsection{Selection criteria}

The Taurus DISCS sample of protoplanetary disks was chosen to assess the impact 
of spectral type or stellar irradiation field on the chemistry in the disk. 
The target systems span the full range of stellar luminosities among the 
T Tauri and Herbig Ae stars associated with the Taurus molecular cloud. 
Table~\ref{tbl:star} lists the stellar properties.
The stellar masses in the sample range between 0.5 and 1.8--2.3 M$_{\odot}$, 
corresponding to luminosities that span almost two orders of magnitude. 
If stellar luminosity is the main driver of the outer disk chemical evolution, 
then this sample should display a range of chemical behaviors. As discussed below there are other sources of radiation that may affect the chemistry as well, especially accretion luminosity and X-rays.

The sample is biased toward disks of large angular extent, since disks smaller 
than a few hundred AU are not spatially resolved by the SMA in the compact 
configuration. The observations are not sensitive to gas inside of 100~AU, which entails that there is a large gap in radii between these millimeter observations and infrared disk chemistry observations that typically probe the inner disk out to a few AUs. The sources were selected from disks previously mapped in CO 
and thus may be biased toward gas-rich disks. Furthermore only disks clearly
isolated from the parent cloud emission are included to reduce confusion and
ensure that the detected molecules reside in the disks. Known harborers of 
organic molecules were favored (DM Tau, LkCa 15 and MWC 480), but the sample 
also contains disks that have only upper limits or that 
have not been investigated for molecular lines other than CO. The focus on a single star 
forming region allows us to probe disks of similar ages and likely similar chemical 
starting points, reducing the sources of chemical variation.

\subsection{Star and Disk Properties affecting Disk Chemistry\label{sec:sample_prop}}

Stellar luminosity, accretion luminosity, X-rays, the interstellar irradiation field, disk geometry, disk gaps and holes, dust settling and dust growth may all affect the chemistry in the disk. The sample characteristics in terms of these properties are discussed in this section.

Observations of high CN and HCN abundances toward protoplanetary disks reveal 
that a chemistry controlled by far-ultraviolet (FUV) or X-ray radiation or 
both must contribute to the observed abundances 
\citep[e.g.][]{vanZadelhoff01,Thi04}. FUV radiation below 2000~\AA~affects the chemistry by directly heating the gas in the disk surface layers, in limiting molecular abundances via photodissociation, increasing the amount of photochemistry products such as CN and liberating frozen species via photodesorption.  The nature of the dominating source of radiation is however unknown. 
The quiescent stellar luminosities in the sample range from 0.25 to 11.5 
L$_{\odot}$, and if quiescent FUV radiation controls the photodissociation 
rate in the disk surface, then there should be a clear trend in the CN/HCN 
ratio between the low-luminosity T Tauri stars and the order of magnitude 
more luminous Herbig Ae stars. However, Kurucz (1993) stellar atmosphere models show evidence that stars with a spectral type later than F do not have significant stellar continuum $<$2000~\AA~above that generated by accretion, and that it is only for A stars that stellar UV becomes more significant than accretion luminosity for the FUV field. This is confirmed by FUV observations toward the T Tauri star TW Hya, which is dominated by line emission generated from accretion shocks \citep{Herczeg02}.

The accretion FUV spectrum is dominated by line emission and especially Ly-$\alpha$ emission, which results in preferential HCN dissociation and therefore boosts the CN/HCN ratio \citep{Bergin03} beyond what is expected from a UV chemistry dominated by continuum radiation. Toward T Tauri stars, the FUV flux is expected to scale with the accretion rate \citep[e.g.][]{Calvet04}.  The accretion rates probably vary over time however, as has been observed toward TW Hya, where \citet{Alencar02} found mass accretion rates between 10$^{-9}$ -- 10$^{-8}$ M$_{\sun}$ yr$^{-1}$ during a year and smaller variations on timescales of days. The measured accretion rate variability among the T Tauri stars in this sample are all within this range and it is unclear whether the average accretion luminosity vary significantly among the sources. In general more massive systems have higher accretion rates \citep{Calvet04} and it can therefore be expected that the FUV flux from accretion will be higher toward the intermediate mass Herbig Ae stars in the sample. As mentioned above,  for the early type stars (e.g. A) the stellar continuum also adds a significant contribution to the FUV radiation field. Thus, even if accretion rather than quiescent luminosity controls the UV chemistry, one would expect to observe a higher CN/HCN ratio toward the more massive stars.

Observations of the FUV field in T Tauri systems find that accretion produces UV fluxes that are a few hundred to a thousand times stronger than the Interstellar Radiation Field (ISRF) at 100 AU \citep{Bergin04}. The ISRF may still be important at large radii, however. 
The external irradiation field is presumed to be constant toward the Taurus sample of sources, i.e. none 
of the disks are close to any O or B stars, but its impact may 
differ between the different classes of sources; the ISRF may play a larger 
role in driving the chemistry for the low luminosity objects, thus acting as a 
chemical equalizer. A fourth possible driver of the disk chemistry is X-rays, which is predicted to be important for the ionization fraction in the disk 
\citep{Markwick02}.  X-rays are mainly attenuated by gas, while continuum UV photons are quickly absorbed by dust  \citep{Glassgold97}. X-rays can therefore penetrate deeper into the disk compared to UV rays, and may be a main driver of both ion chemistry and molecular dissociation. This may result in observable differences in the molecular distribution if the chemistry is driven by X-rays instead of UV radiation. For individual objects X-ray measurements are notoriously variable and assessing their impact observationally may be possible only through monitoring of the X-ray flux and the chemical variation toward a single system \citep{Glassgold05}. On average the X-ray fluxes seem higher toward T Tauri stars compared to Herbig Ae stars and thus ion-driven chemical reactions may be faster in disks around T Tauri stars \citep{Gudel09}.

Regardless of whether the quiescent luminosity 
of the central star is the main source of energetic radiation, it may still
control the disk temperature and thus the temperature sensitive chemistry, 
e.g. deuterium fractionation efficiency probed by the DCO$^+$/HCO$^+$ ratio.
The stellar continuum photons are expected to be the primary agent for heating grains in the disk, particularly toward the midplane, setting the overall reservoir of warm or cold grains in the outer disk to which the SMA is most sensitive. Both the level of gas phase depletion and the chemistry dependent on CO depletion may then be regulated by the stellar continuum flux. For example a tracer of CO freeze-out such as the N$_2$H$^+$/HCO$^+$ ratio is expected to be higher toward low-luminosity systems \citep{Bergin02}.

The disk structure and dust properties determine how much of the stellar 
radiation is intercepted by the disk and the penetration depth of 
that radiation. These disk properties may be equally important to the 
the strength of the radiation field for the chemical evolution in the disk. Table \ref{tbl:disk} 
list the sample disk characteristics from previous CO and continuum 
observations and modeling. Position angles and inclinations do not affect the 
chemistry intrinsically, but the disk inclination affects which parts of 
the disk are observed, and thus our view of the chemistry. The sizes of the disks are described in terms of CO gas disk radii and range
from 200 to 890~AU. Disk masses are less well constrained, since they are 
derived from dust emission assuming a dust-to-gas ratio. In most studies the canonical interstellar dust-to-gas ratios of 1 to 
100 is used. The actual dust-to-gas ratio may be different because of dust coagulation and photoevaporation, and also variable among the sources. Still it seems clear that two of the disks, AA Tau and CQ Tau, 
are significantly less massive than LkCa 15, GM Aur and MWC 480, while DM Tau 
falls in between (Table \ref{tbl:disk}). This difference in disk mass may affect their relative 
abundances of different species, since more massive disks are expected to 
contain more cold material per unit of incoming irradiation. 
 
Three of the sample disks are so-called transition and pre-transition disks (DM Tau, GM Aur and 
LkCa 15) with large inner holes or gaps \citep{Calvet04, Espaillat07,Dutrey08}. From a {\it Spitzer} survey of disk chemistry many transition disks, including GM Aur and LkCa 15 have CO gas in the disk hole \citep{Salyk09}. They do however lack emission from HCN and C$_2$H$_2$ transitions in the {\it inner} disk that are strong toward classical T Tauri stars \citep{Pascucci09}.     It is unclear whether this is a chemistry or gas-mass effect. It is also unknown whether the chemistry in 
the outer regions of these transition disks differ significantly from classical 
T Tauri disks, although it is curious that these systems are the most chemically
rich in terms of the outer disk seen to date. This may be explained by increased radiation fluxes; more stellar radiation may reach the outer disk the larger the hole. In addition, large holes may be a tracer of overall grain growth and thus of increased UV penetration depth in both the inner and outer disk.

Grain properties can be traced by the millimeter slope of the Spectral Energy 
Distributions (SEDs), parameterized by the power-law
index of the opacity spectrum, $\beta$, and this is the last disk 
characteristic listed in Table \ref{tbl:disk}. The index $\beta$ is predicted to decrease with grain growth and the disks in the survey all 
have $\beta$ below the expected 1.7 for interstellar grains \citep{Andrews07}.  The $\beta$ estimates 
are however different between different studies, which makes it difficult to 
conclude on an order of dust coagulation among the disks, but AA Tau seems to 
have a significantly lower $\beta$ compared to the other T Tauri disks. Most of the sources have also been observed by Spitzer in studies that constrain the grain properties in the inner disk \citep{Furlan09}. These measurements do not provide any straightforward constraints on the grain properties in the outer disk probed by the SMA, however. 

Finally the disk structure, e.g. the amount of flaring, affects the amount of 
stellar light intercepted by the disk. The disk structure, parameterized by the dust scale height, can be constrained by 
modeling the SED, though degeneracy is a problem \citep{Chiang01}. As the dust settles towards the midplane the dust scale height decreases compared to the gas scale height. Five of the sample disk SEDs were modeled by \citet{Chiang01}, who concluded that MWC 480 and LkCa 15 are more settled than CQ Tau, GM Aur and AA Tau, i.e. in the disks of MWC 480 and LkCa 15 the upper disk layers are significantly depleted in micron-sized dust grains. This has been confirmed in more recent studies that find significantly less settling toward GM Aur compared to LkCa 15 \citep{Espaillat07, Hughes09}.

\section{Observations\label{sec:obs}}

\subsection{Spectral Setups}
Two frequency setups per source were selected to cover 4 to 8 spectral lines 
in each setup, at the same time providing continuum observations.
The targeted molecules are DCO$^+$ and DCN (probing deuterium chemistry), 
CN and HCN (probing photochemistry), HCO$^+$ and N$_2$H$^+$ (ions), H$_2$CO 
and CH$_3$OH (potential grain chemistry products), $c$-C$_3$H$_2$ (carbon-chain chemistry) and CO (the disk kinematic tracer). 
Tables ~\ref{tbl:setup1} and ~\ref{tbl:setup2} summarizes the two spectral 
setups, centered at 1.1 mm and 1.4 mm, respectively. 

The SMA correlator covers $\sim4$~GHz bandwidth in each of the two sidebands 
using two Intermediate Frequency (IF) bands with widths of 1.968 GHz.
The first IF band is centered at 5 GHz, and the second IF band is centered 
at 7 GHz. Each band is divided into 24 slightly overlapping ``chunks'' of 
104 MHz width, which can have different spectral resolution. 
For each spectral setting, the correlator was configured to increase the 
spectral resolution on the key species with 128--256 channels per chunk, with 
the exception of one H$_2$CO line observed with 64 channels. Chunks containing weaker lines were then binned to obtain higher signal to noise, while still recovering sufficient kinetic information. The remaining chunks were covered by 64 channels each and used to measure 
the continuum. The continuum visibilities in each sideband and each IF band 
were generated by averaging the central 82 MHz in all line-free chunks.

\subsection{Data Acquisition and Calibration}

The six disks were observed from 2009 November through December with the 
compact configuration of the Submillimeter Array (SMA) interferometer 
(Ho et al. 2004) at Mauna Kea, Hawaii. LkCa~15 was also observed on 2010 March 22nd with the subcompact configuration
for more short-spacing data to improve the signal-to-noise of deuterated
species detection. For each observation, at least six 
of the eight 6~m SMA antennas were available, spanning baselines of 16--77~m 
(Table ~\ref{tbl:obs}).  
The observing sequence interleaved disk targets and two quasars in an 
alternating pattern. Depending on their proximity to the disk targets
and fluxes at the time of the observations, a group of three quasars was used 
as gain calibrators: 3C 111, J0510+180 and 3C 120. The observing conditions 
were generally very good, 
with $\tau_{\rm 225{\:GHz}}\sim0.04-0.1$ and stable atmospheric phase.

The data were edited and calibrated with the IDL-based MIR software 
package\footnote{http://www.cfa.harvard.edu/$\sim$cqi/mircook.html}.
The bandpass response was calibrated with observations of Uranus and
the bright quasars available (3C 454.3, 3C 273, 3C 84).
Observations of Uranus provided the absolute scale for the calibration of 
flux densities for all compact tracks and Vesta for the one LkCa 15 sub-compact track. The typical systematic uncertainty in the absolute flux 
scale is $\sim$10\%. Continuum and spectral line images were generated 
and CLEANed using MIRIAD. MIRIAD was also used to calculate synthesized beam sizes and position angles, which are listed in Table \ref{tbl:beam} for each setting and source.

\section{Results and simple analysis}\label{sec:res}

The first section presents the extracted spectra for all molecules detected toward at least one disk, channel maps toward DM Tau, which contain some of the strongest detections of the weaker lines in the survey and finally disk images and moment maps for strong 
molecular lines toward all sources
($\S$\ref{ssec:overview}). The subsequent sections provide more details 
on the observed CN/HCN ratios, ions and deuterated molecules and their 
variation among the surveyed disks. 

In general, we do not attempt to estimate column densities of molecules, but rather present the data in terms of integrated fluxes and flux ratios. The line fluxes and column densities are of course related, but the fluxes also depend on excitation temperatures and opacities. Determining molecular column densities therefore requires knowledge of the disk structure and the spatial distribution of molecules. Even for optically thin lines, estimated column densities may be off by an order of magnitude or more if the wrong emission regions/temperatures are adopted. For optically thick lines the estimates become even more uncertain and single dish studies including $^{13}$CO and H$^{13}$CO$^+$ have shown that the  emission from $^{12}$CO and H$^{12}$CO$^+$ is mostly optically thick \citep{Dartois03,Thi04}.   A proper derivation of column densities therefore requires detailed chemical modeling
 of the disks, which we anticipate for a future DISCS publication. An exception from only reporting fluxes is made for the deuterium fractionation, where it is useful to derive abundance ratio limits assuming the same emission regions of molecules and their deuterated equivalents, to compare with previous studies. We also estimate the HCO$^+$/CO ratio toward CQ Tau for a rough consistency check with previous observations.

\subsection{Overview of Taurus disk chemistry\label{ssec:overview}}

\begin{figure}
\epsscale{1.0}
\plotone{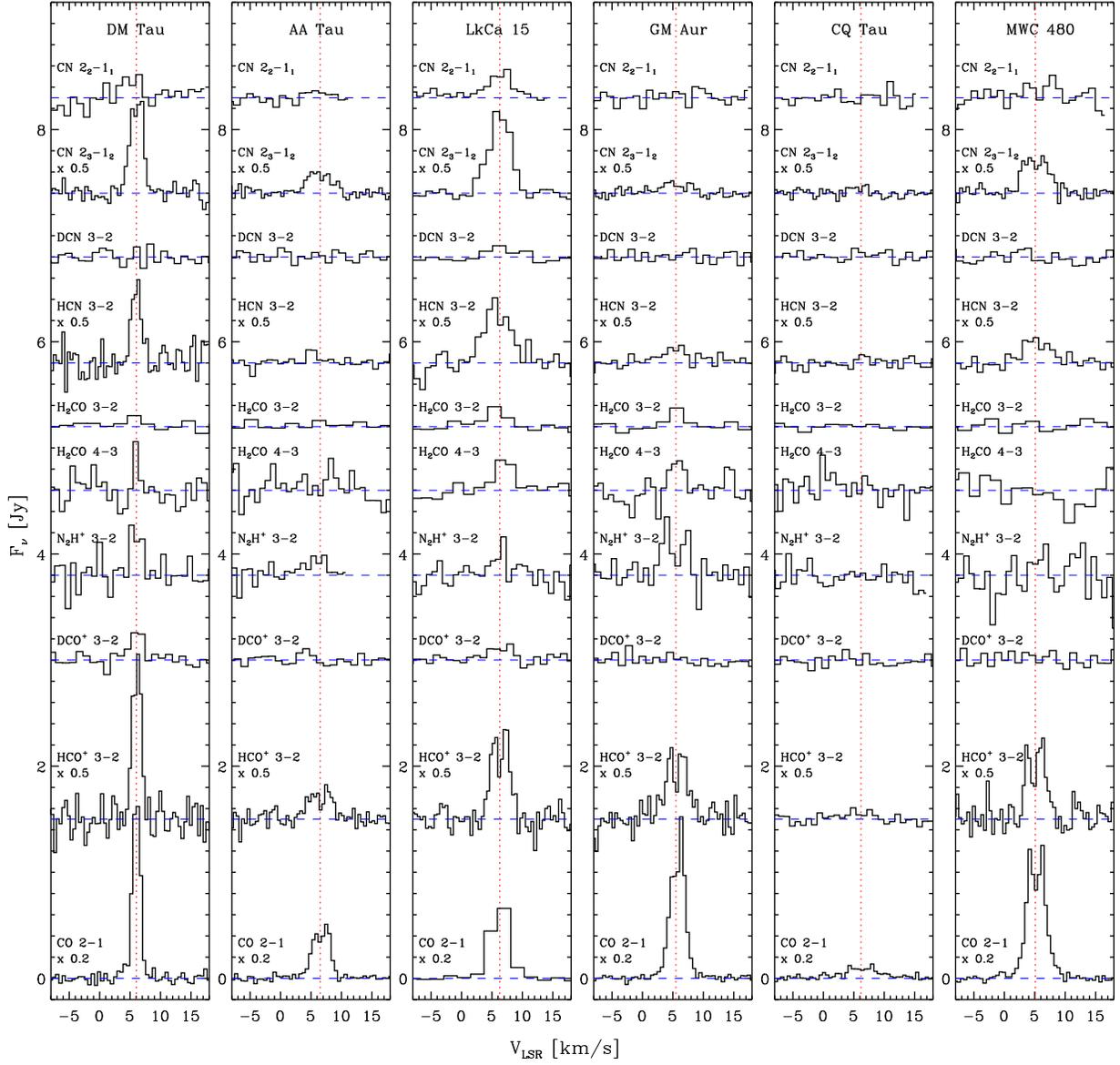}
\caption{Spatially integrated spectra from DM Tau, AA Tau, LkCa 15 GM Aur, CQ Tau and MWC 480, displaying the chemical variation between the different disks. The line profiles of CO, HCO$^+$, HCN and one of the CN lines have been scaled down for visibility. \label{fig:spec}}
\end{figure}

Figure \ref{fig:spec} shows the extracted spectra of the targeted molecules toward DM Tau, AA Tau, LkCa 15, GM Aur, CQ Tau and MWC 480. The spectra are also tabulated in Table \ref{tbl:spec}. The four spectral lines, CO 2--1, HCO$^+$ 3--2, HCN 3--2 and CN 3--2, expected 
to be strongest from previous studies were detected in all six disks except for HCN toward CQ Tau, with an order of magnitude variation in integrated fluxes between the different disks. Spectral lines from 
N$_2$H$^+$ and H$_2$CO were detected toward three disks, DM Tau, LkCa 15 and 
GM Aur, with tentative detections toward AA Tau. Lines from the two deuterated 
molecules, DCO$^+$ and DCN, were detected toward LkCa 15 and the DCO$^+$ 3--2 
line was also detected toward DM Tau. The spectral lines from CH$_3$OH and 
c-C$_3$H$_2$ were not detected toward any of the disks.

\begin{figure}
\epsscale{0.9}
\plotone{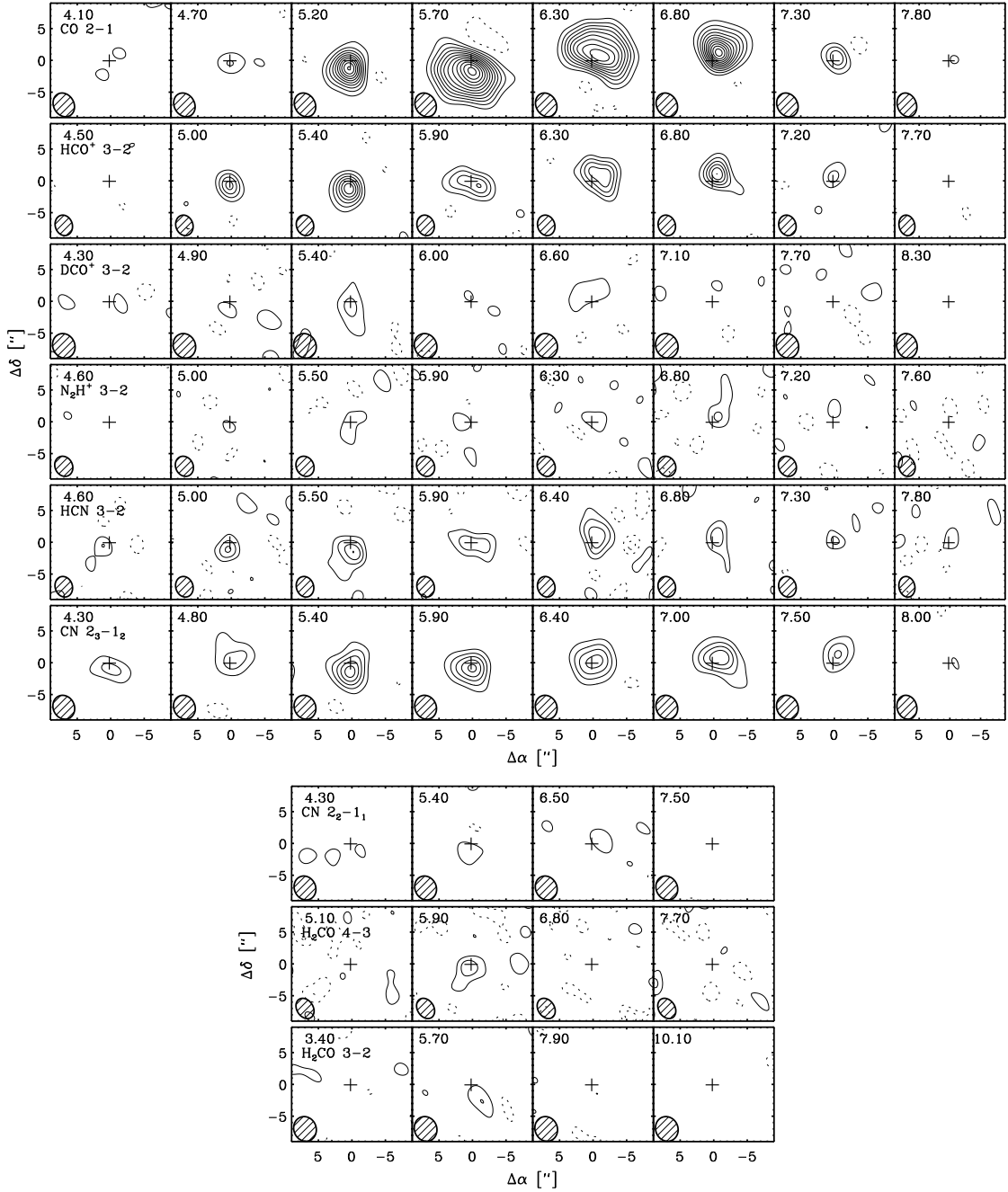}
\caption{{\small Velocity channel maps for all detected molecules toward DM Tau . The first contours as well as the contour step values are $\sim$ 2-$\sigma$, or 0.15 Jy beam$^{-1}$ for CO and HCO$^+$, 0.12 Jy beam$^{-1}$ for N$_2$H$^+$ and HCN, 0.07 Jy beam$^{-1}$ for DCO$^+$, H$_2$CO 4--3 and the CN lines, and 0.04 Jy beam$^{-1}$ for H$_2$CO 3--2. he cross indicates the continuum (stellar) position. Note the different velocity resolution for the bottom four transitions.} \label{fig:dmtau}}
\end{figure}

Figure \ref{fig:dmtau} shows the velocity channel maps for all detected 
spectral lines toward DM Tau. The peak intensities vary between  $\sim$9 Jy beam$^{-1}$ 
for CO and $\sim$0.1 Jy beam$^{-1}$  for the weaker H$_2$CO line. All species, except for H$_2$CO, are apparent in 
multiple velocity channel maps, and the stronger detections clearly follow 
the velocity pattern indicative of a disk in Keplerian rotation. The relative intensity of different species in different velocity channels varies significantly, indicative of differences in emission regions between different lines.  

\begin{figure}
\epsscale{1.0}
\plotone{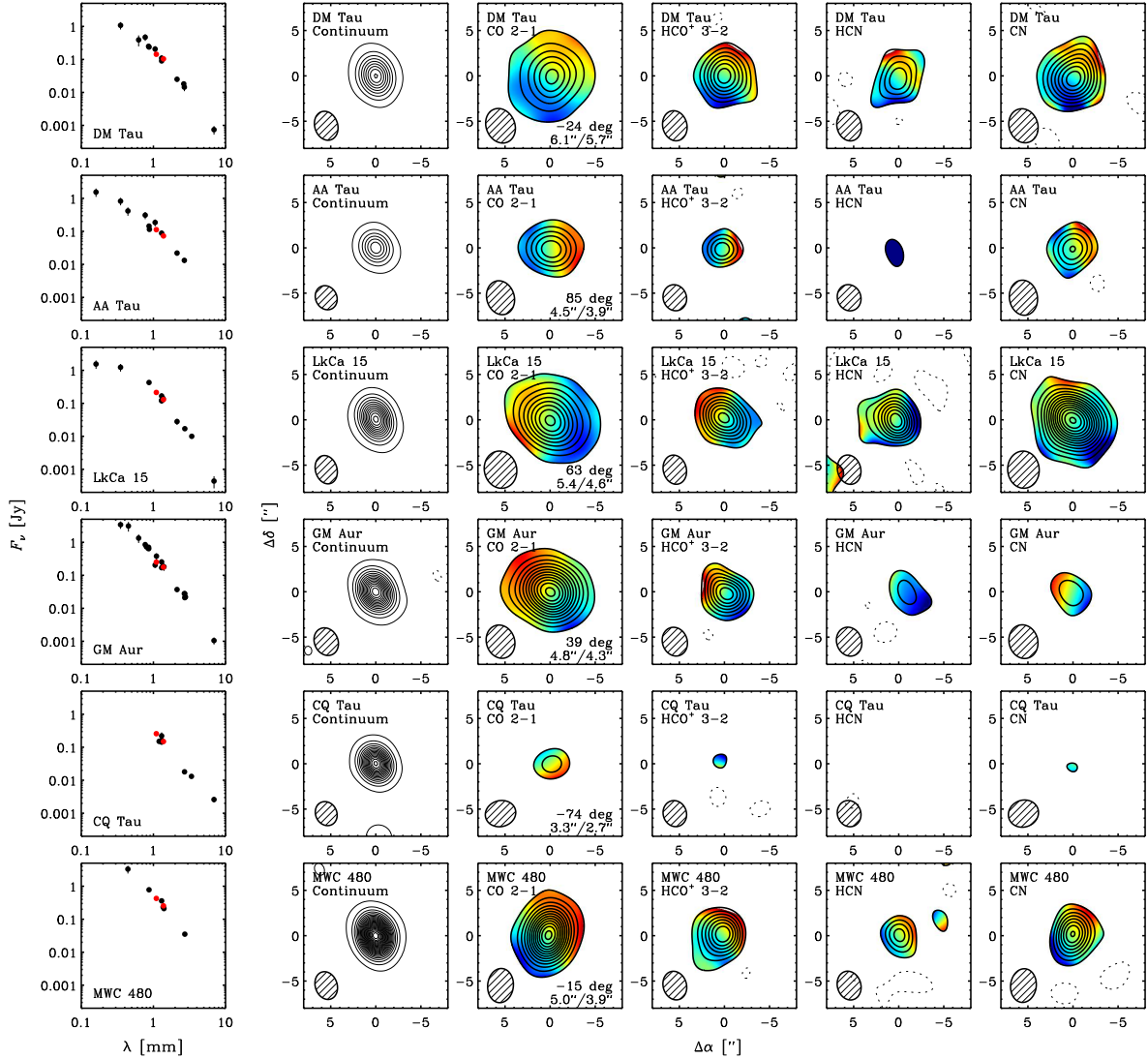}
\caption{Literature spectral energy distributions with our continuum points superimposed in red (col 1), 267 GHz continuum maps (col 2) and zero- and first moment maps of CO 2-1, HCO$^+$ 3-2, HCN 3-2 and CN 2-1 (col 3-6). In the moment maps the contours show the integrated intensity in steps of 0.8, 0.5, 0.32 and 0.32 Jy beam$^{-1}$ km s$^{-1}$ for CO, HCO$^+$, HCN and CN respectively. The velocity gradient is defined over the same range for all molecules toward each source. The derived position angles and CO FHWM major and minor axes derived from fitting ellipses to the CO maps are shown in the lower right corner of each CO panel.  \label{fig:maps}}
\end{figure}

Fig. \ref{fig:maps} shows integrated intensity and first moment maps derived 
from the channel maps for CO, HCO$^+$, HCN and CN for each disk.
In addition, the first column in Figure \ref{fig:maps} shows millimeter
continuum SEDs compiled from the literature \footnote{\citet{Acke04,Adams90,Andrews05, Beckwith90, Beckwith91,Chapillon08,Dutrey96,Dutrey98,Duvert00,Guilloteau98,Hamidouche06,Hughes09,Isella09,Kitamura02,Koerner93,Looney00,Mannings94,Mannings97,Mannings00,Natta01,Osterloh95,Pietu06,Rodmann06,Testi01,Weintraub89}} together with new measurements
at 218 GHz and 267 GHz (Table  \ref{tbl:int}), which show good agreement for all of the sources.
The dust continuum flux densities at $\sim$267 (and 218) GHz vary by about a factor of 
four among these sources, with MWC 480 the strongest at 430 mJy and
AA Tau the weakest at 110 mJy. Note that there is not a one-to-one  
correspondence between the strength of the 218 or 267 GHz dust continuum emission
and the CO 2-1 line emission.

The first-moment maps show the Keplerian rotation pattern of the disks, most 
clearly in the CO 2-1 and HCO$^+$ 3-2 emission. The rotational velocity 
pattern is also present to some extent in HCN and CN emission, which suffer 
from lower signal-to-noise and, for CN, the blending of the 
spectroscopic triplet. 

Line spectra are extracted from the channel maps using elliptical masks 
produced by fitting a Gaussian profile to the CO integrated intensity maps 
toward each source to obtain major and minor axes and positions angles (listed in Fig. \ref{fig:maps}). 
The size of the mask is scaled for each line, to optimize the signal-to-noise without losing any significant emission, 
such that the major and minor axes are between 2-$\sigma$ and 1-$\sigma$ of 
the Gaussian fitted to the CO emission -- a 2-$\sigma$ radius corresponds to 
$\sim$1.7 $\times$ the radius at full-width-half-maximum (FWHM). All masks are large enough to not cut out any emission, i.e. the chosen masks results in no significant decrease in integrated line intensity compared to integrating over the full CO disk. Applying these 
masks ensures that only disk emission and minimal noise are included in the 
spectra.  The derived CO full-width-half-maxima agree reasonably well with the disk
sizes from the literature compiled in Table \ref{tbl:disk}, i.e. all semi-major 
axes are within $1''$ of the previously observed or derived CO radii, where 
resolved data exist. In addition the position angles agree within 
10$\degree$ except for the barely resolved disk of CQ Tau. The integrated spectra from CQ Tau are however not significantly affected by the choice of mask shape. 

The resulting spectra in Fig. \ref{fig:spec} are used to derive the total intensities listed in   Table \ref{tbl:int}. The 2-$\sigma$ upper limits are calculated from the rms when the spectra are binned to a spectral resolution of 3.3--4.3 km s$^{-1}$, the minimum resolution for resolving any lines, multiplied by the full width half maximum of the CO 2-1 line toward each source. This approximation is supported by the similarity in the line widths for different transitions toward the same disk in Fig. \ref{fig:spec}. Because of variable observing conditions and integration times, the flux upper limits vary between 0.09 and  0.78 Jy km s$^{-1}$ per beam. Most upper limits are however lower compared to detected fluxes toward other sources. Specifically the non-detections toward MWC 480 of DCO$^+$, N$_2$H$^+$ and one of the H$_2$CO lines are up to a factor of two lower than the detected fluxes toward the T Tauri stars DM Tau, GM Aur and LkCa 15.

\subsection{CN/HCN flux ratios}

As discussed above, deriving molecular abundances from disk spectra is complicated by the uncertainties in disk structures and the accompanying uncertainties in the emission conditions of the different molecules. Line flux ratios can however be used as a proxy for abundance ratios when the emission is optically thin and the upper energy levels are similar, canceling out the temperature effect, if the same emission region for the two molecules can be assumed. Models suggest that the emission regions are different for CN and HCN, however \citep{Jonkheid07}. It is still informative to compare the flux ratios between different sources if it can be assumed that the relative emission regions of the CN and HCN is stable between different disks, i.e. that CN is always present in a warmer layer closer to the surface compared to HCN. Then changes in the CN/HCN flux ratio can still be used to trace changes in abundance ratios if the emission is optically thin.

\citet{Thi04} suggested that HCN and CN line emission from protoplanetary 
disks may be somewhat optically thick from comparisons of HCN and H$^{13}$CN 
line intensities toward LkCa 15.  This survey does not currently include any 
rare isotopologues, but the CN optical depth can be estimated from the relative 
integrated intensities of the 2$_3$-1$_2$ triplet transition and the $\sim$10 times weaker 
2$_2$-1$_1$ singlet transition using the line strengths
from CDMS \citep[http://www.astro.uni-koeln.de/cdms/catalog and][]{Muller01} 
at any of the reported temperatures between 9 and 300~K (the transitions have 
the same excitation energy). Toward DM Tau and AA Tau the observed CN line 
ratios are consistent with optically thin emission. Toward LkCa 15 and MWC 480, the emission from the CN triplet underestimates the CN abundance by  factors of 1.3 and 1.6, respectively, indicative of somewhat optically thick 
CN triplet emission.

\begin{figure}
\epsscale{1.0}
\plotone{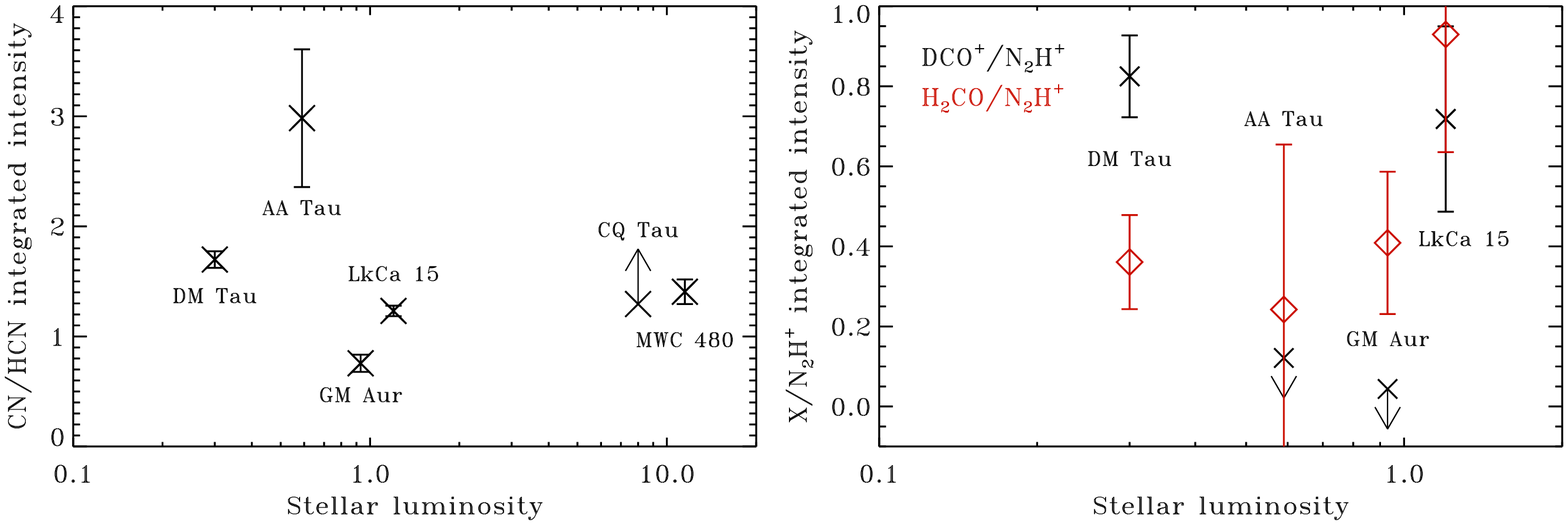}
\caption{Integrated intensity line ratios and upper and lower limits for CN/HCN and DCO$^+$/N$_2$H$^+$ and H$_2$CO/N$_2$H$^+$. \label{fig:ratios}}
\end{figure}

Despite the complications introduced by modest 
optical depth,
changes in CN 2$_3$-1$_2$ / HCN 3--2 ratios larger than a factor of 2, 
assuming similar levels of optical depth for HCN emission, are expected to 
trace real variations in the chemistry as discussed above. 
The absolute CN and HCN integrated intensities range from 0.2 Jy km s$^{-1}$ 
toward CQ Tau to 5.5 Jy km s$^{-1}$ toward LkCa 15. 
Figure \ref{fig:ratios} shows that the variation in CN/HCN line intensities 
is smaller -- all sources have ratios of 0.8--1.6 except for AA Tau, which has 
a CN/HCN emission ratio of 2.9. AA Tau is then the only significant outlier 
in terms of CN/HCN flux ratios.

\subsection{Ions: HCO$^+$ (DCO$^+$) and N$_2$H$^+$}

HCO$^+$ is often used as a tracer of gas ionization and thus of high energy 
radiation in disks. The high optical depth of HCO$^+$ and the lack of rare 
isotopologues of CO prevent such an analysis at present, except for toward 
CQ Tau where the CO 2--1 emission has been estimated to be optically thin 
\citep{Chapillon08}. The ratio of column densities for species X and Y, for 
optically thin, resolved emission, can be calculated from

\begin{equation}
\label{eq:ratio}
\frac{N_{\rm X}}{N_{\rm Y}} = \frac{\int T_{\rm mb}^{\rm X} d\nu}{\int T_{\rm mb}^{\rm Y} d\nu} \times \frac{Q_{\rm rot}^{\rm X}(T)}{Q_{\rm rot}^{\rm Y}(T)} \times \frac{e^{E_{\rm u}^{\rm X}/T_{\rm ex}}}{e^{E_{\rm u}^{\rm Y}/T_{\rm ex}}} \times \frac{\nu_{\rm Y} S_{\rm Y}\mu^2_{\rm Y}}{\nu_{\rm X} S_{\rm X}\mu^2_{\rm X}},
\end{equation}

\noindent where $N$ is the column density, $\int T_{\rm mb} d\nu$ is the integrated line 
emission in K km s$^{-1}$,  $Q_{\rm rot}(T)$ the temperature dependent 
partition function, $E_{\rm u}$ the energy of the upper level in K, 
$T_{\rm ex}$ the excitation temperature in K and $S_{\rm Y}\mu^2$ are the line 
strength and dipole moment \citep[e.g.][]{Thi04}. Toward the same source, the 
integrated line flux in Jy and line intensity in K are related by 
$T_{\rm mb}[{\rm K}]\varpropto F[{\rm Jy}]\times\lambda^2[{\rm mm^2}]$. 
Using partition functions, level energies and line strengths from CDMS and 
assuming the same excitation conditions for HCO$^+$ 3--2 and CO 2--1 the 
[HCO$^+$]/[CO] ratio is $1.0-1.3\times10^{-4}$ toward CQ Tau for excitation temperatures of 
18--75~K.

The N$_2$H$^+$ observations toward DM Tau, LkCa 15 and GM Aur (and tentatively
 toward AA Tau) provide unambiguous detections of this species in protoplanetary
 disks, confirming previous claims by 
\citet{Dutrey07}. Since N$_2$H$^+$ and DCO$^+$ both potentially trace the chemistry further toward the midplane compared to the more abundant molecules, their ratio provides important constraints on the cold chemistry in disks. The DCO$^+$/N$_2$H$^+$ line intensity ratio ranges from $<$0.1 toward GM Aur to 0.8 toward DM Tau (Fig. \ref{fig:ratios}). In contrast, there is no significant variation in the H$_2$CO/N$_2$H$^+$ ratio between the sources. Assuming that these molecules always reside in the colder regions of the disks, these differences in flux ratios suggest relative abundance variations of N$_2$H$^+$ and DCO$^+$ of an order of magnitude between the different sources. 
 
\subsection{Deuteration: DCO$^+$/HCO$^+$ and DCN/HCN}

Because of the high optical depth of the HCO$^+$ line emission, the DCO$^+$/HCO$^+$ line intensity ratio can only be used to derive upper limits on the average HCO$^+$ deuteration fraction in the disk. The analysis is further complicated by evidence of different emission regions of DCO$^+$ and HCO$^+$ \citep{Qi08}. Assuming, however, the same emission region of DCO$^+$ and HCO$^+$ and optically thin emission for both ions, the upper limits on the deuteration fraction is calculated using Eq. \ref{eq:ratio} to vary between 0.32 toward DM Tau, 0.18 toward LkCa 15 and $<$0.07 toward GM Aur. An excitation temperature of 19 K is assumed, but the ratios are only marginally affected by the excitation temperature between 10 and 50~K.  The variable ratios hint at differences in deuteration fractionation between different sources, especially since GM Aur has both lowest DCO$^+$/HCO$^+$ and DCO$^+$/N$_2$H$^+$ ratios, but radiative transfer modeling is needed to confirm this result. 

DCN is only detected toward LkCa 15. The detection is at the $>$5-$\sigma$ level, and the moment maps shows an almost perfectly aligned velocity field compared to CO and HCO$^+$ and the detection appears secure. From the CN analysis above and previous single-dish observations, HCN line emission is expected to be much less optically thick than HCO$^+$ emission and the DCN/HCN ratio should provide stricter limits on the deuteration level in the disk. Assuming optically thin emission, the same emission region for DCN and HCN and an excitation temperature of $\sim$40~K, the upper limit on average deuteration in the disk around LkCa 15 is 0.06, a factor of three lower than the estimate from DCO$^+$/HCO$^+$.

\section{Discussion} \label{sec:disc}

\subsection{Detection rates and comparison with previous studies}

The reported images and spectra were acquired with only 3--7 hours on source integration, with the shorter time spent in the 1.1 mm setting. The detection rate of N$_2$H$^+$ and H$_2$CO in the 1.1 mm setting is then quite remarkable and can be attributed to the advantages of targeting higher $J$ lines when probing disks. For comparison, N$_2$H$^+$ was previously detected toward DM Tau and LkCa 15 through its 1--0 line using the Plateau de Bure Interferometer, with peak fluxes $<$0.02 Jy \citep{Dutrey07}, an order of magnitude or more lower than the $3-2$ line peak fluxes reported here. The 1-0 line in LkCa 15 was also detected by \citet{Qi03} using
the Owens Valley Radio Observatory Millimeter Array with integrated intensity four times larger than that with the PdBI. 

The agreement with previous single-dish observations is generally good. All targeted molecular lines that have been previously reported in the 
single dish studies of DM Tau, LkCa 15 and MWC 480 are also detected with the 
SMA \citep{Dutrey97,Thi04, Guilloteau06}. Where the same lines have been 
studied, most integrated intensities agree. HCN $J=3-2$ toward DM Tau is an exception, 
where the reported upper limit in \citet{Dutrey97} is a factor of two below 
the intensity observed with the SMA.

Using their detections, \citet{Dutrey07} derived [N$_2$H$^+$]/[HCO$^+$] ratios of 0.02--0.03 for DM Tau and LkCa 15 by fitting the line emission to disk models. Without such modeling we can only derive upper limits on the [N$_2$H$^+$]/[HCO$^+$] of 0.13--0.19 for the two disks because of the HCO$^+$ line optical depth. Considering that the HCO$^+$ abundance may be underestimated by up to an order of magnitude, the two data sets are consistent. Within the same observational program \citet{Chapillon08} searched for CO 2--1 and HCO$^+$ 1--0 emission toward CQ Tau and used the data to derive an upper limit on the [HCO$^+$]/[CO] abundance ratio. Assuming the same CO and HCO$^+$ distribution and excitation conditions and optically thin CO emission they find [CO]/[HCO$^+$]$>$4$\times10^3$. This is consistent with the abundance ratio of 10$^4$ reported above, which is calculated making similar assumptions, but without the detailed modeling in \citet{Chapillon08}. 

CQ Tau is by far the most chemically poor of the investigated disks. It is interesting that despite the low abundances, the chemistry appears 'normal', the ratios of the integrated intensities toward MWC 480 and CQ Tau are the same within a factor of two, including the CN/HCN emission ratio. The only difference is that overall the gas toward CQ Tau is probably richer in CN and HCN with respect to CO, taking into account the large optical depth of the CO emission toward MWC 480 \citep{Thi04}, as might be expected for a smaller disk, completely exposed to UV radiation.

The upper limits on the [DCO$^+$]/[HCO$^+$] abundance ratio of $<$0.07--0.32 found toward the T Tauri systems in DISCS are consistent with the ratio of 0.035--0.05 observed toward TW Hydrae \citep{vanDishoeck03,Qi08}. The better constrained [DCN]/[HCN] ratio of 0.06 toward LkCa 15 is also consistent with the value of 0.02--0.05 toward TW Hydrae. While this ratio may be overestimated by a factor of a few, high levels of deuterium fractionation seems common toward T Tauri systems. 

As found in single dish studies, the CN/HCN line intensity ratios toward the Taurus disks are high compared to interstellar clouds and cores. The line ratios all fall within the range of measurements toward other disks, where the total integrated flux ratio of CN/HCN is $\sim$1--5 \citep[see][for a compilation]{Kastner08}. A more quantitative comparison with previous observations is difficult without detailed modeling because different studies observed different transitions of CN and HCN.

\subsection{T Tauri vs. Herbig Ae stars}

In agreement with previous studies, we find that the disks surrounding T Tauri stars are more chemically rich in species with strong mm-transitions compared to disks around Herbig Ae stars \citep{Chapillon08,Schreyer08,Henning10}. The observed chemical poverty in the outer disks of Herbig Ae stars has been attributed to the more intense UV field around Herbig Ae stars compared to T Tauri stars, which may efficiently photodissociate most targeted molecules. In this sample, the most obvious difference between T Tauri and Herbig Ae stars is the lack of the cold chemistry tracers N$_2$H$^+$, DCO$^+$, DCN and H$_2$CO toward CQ Tau and MWC 480, while they are detected in 3/4, 2/4, 1/4 and 3/4 of the T Tauri systems. The upper limits toward CQ Tau are less informative because of its weak CO emission and observations toward more Herbig Ae stars are required to confirm that this difference between disks around low and medium mass stars is general. 

In contrast the CN and HCN emission is similar toward the lowest and highest luminosity stars, DM Tau and MWC 480, in the sample. CN and HCN emission are modeled to originate mainly from the outer layers of the disk \citep{Willacy07} and this chemistry thus seems equally active toward low- and intermediate-mass pre-main sequence stars.  

\subsection{CN and HCN}

CN is a photodissociation product of HCN and the CN/HCN ratio has been put forward to trace several different aspects of the UV field. The CN/HCN ratio is proposed to increase with the strength of the UV field \citep{vanZadelhoff03}, and it will be further enhanced if the UV radiation is dominated by line emission from accretion, since HCN is dissociated by Ly-$\alpha$ photons while CN is not \citep{Bergin03}. Dust settling or coagulation allows radiation to penetrate deeper into the disk, which is also predicted to enhance the CN/HCN ratio \citep{Jonkheid07}.  

The quiescent UV luminosity increases with stellar mass. There is however no 
visible trend in the emission ratio of CN/HCN with spectral type. 
In fact, all CN/HCN ratios are the same within a factor of two, except toward AA Tau, 
which has a factor of a few higher intensity ratio. This suggests that the 
CN/HCN ratio is not set by the stellar luminosity though there are complications in comparing CN/HCN ratios toward disks around low- and intermediate-mass stars because of potentially different excitation conditions for HCN in the two sets of disks \citep{Thi04}. Within this sample the CN/HCN ratio also does not trace accretion luminosity; AA Tau has a comparable accretion rate to LkCa 15 and among the sources with comparable CN/HCN ratios the accretion rate varies by an order of magnitude. AA Tau is reported to have a lower power-law
index of the opacity spectrum, $\beta$, compared to the other disks, indicative of dust growth and it may be the dust properties rather than the stellar or accretion luminosities govern the importance of photochemistry in disks. A larger sample that spans a wider range of accretion rates and dust properties is clearly required to give a more definitive answer.

An additional complication is that the high CN/HCN ratio toward AA Tau may be a geometric effect. Compared to the other disks, AA Tau is almost edge-on \citep{Menard03}, which may result in preferential probing of the disk atmosphere compared to less inclined disks. Disk chemistry models (Fogel et al. submitted to ApJ) show that CN mainly emits from the disk surface, while HCN emission originates further into the molecular disk layer and the more inclined disk may offer a viewing angle that is biased toward CN emission. To estimate the effect of disk inclination on CN/HCN flux variations then requires a combination of chemical modeling and radiative transfer models. 

In terms of absolute flux intensities, the weak CN and HCN emission toward GM Aur compared to LkCa 15 and DM Tau stands out. The difference between GM Aur and LkCa 15 may be due to the higher accretion rate and intenser FUV field toward LkCa 15. The difference between DM Tau and GM Aur is however difficult to explain in terms of UV flux, since DM Tau is a weaker accretor than GM Aur. There is some evidence for significantly more dust settling toward LkCa 15 and DM Tau compared to GM Aur \citep{Chiang01,Espaillat07,Hughes09}. This may expose more of the gas in the LkCa 15 and DM Tau disks to high-energy radiation, enhancing the photoproduction  of CN and HCN as well as the ion chemistry deeper in toward the disk midplane. 

\subsection{Cold chemistry tracers}

Lower abundances of DCO$^+$, DCN, N$_2$H$^+$ and H$_2$CO toward more luminous stars are qualitatively consistent with our current chemical understanding. 
DCO$^+$ forms efficiently from gas phase reactions with H$_2$D$^+$, which is 
only enhanced at low temperatures \citep{Roberts00,Willacy07} and should be enhanced
toward colder disks. Among the T Tauri stars the brightest DCO$^+$ emission is observed toward the disk around the least luminous star, DM Tau, consistent with a higher degree of deuterium fractionation around colder stars. The difference 
in DCO$^+$ line flux around GM Aur and LkCa 15 is more difficult to explain. GM Aur has a more massive dust disk than LkCa 15 and the two stars have similar luminosities. Naively GM Aur should then be surrounded by at least as much cold disk material as LkCa 15. Instead, the upper limit on the DCO$^+$ flux is a factor of three lower toward GM Aur compared to LkCa 15. There is thus no one-to-one correlation between the ratio of  disk dust mass over quiescent stellar luminosity and DCO$^+$ column densities.

N$_2$H$^+$ forms from protonation of N$_2$ by H$_3^+$ and is mainly destroyed by reactions with CO \citep{Bergin02}. Abundant N$_2$H$^+$ is therefore only expected where CO is depleted onto grains toward the disk midplane. N$_2$ freezes onto grains a few degrees below CO \citep{Oberg05} and in cold disks the N$_2$H$^+$ abundance should peak in a narrow region where the temperature is between the N$_2$ sublimation temperature of $\sim$16~K and the CO sublimation temperature of $\sim$19~K. As long as a cold region exists in the disk, the N$_2$H$^+$ abundances may be quite independent of the total amount of cold disk material. The observations are consistent with this disk abundance structure; within the T Tauri sample the N$_2$H$^+$ emission only varies by a factor of two, increasing slightly with increasing disk mass. 

The variation in DCO$^+$/N$_2$H$^+$ flux ratios over the sample suggest that while both molecules trace a cold chemistry, their dependences on the physical environment is considerably different. The $>$8 times higher DCO$^+$/N$_2$H$^+$ flux ratios toward DM Tau and LkCa 15 compared to GM Aur may be related to the 5--6 times higher fluxes of CN and HCN toward DM Tau and LkCa 15 compared to GM Aur. This would suggest that both ratios  depend on the amount of dust settling and that DCO$^+$ trace a cold radiation driven chemistry. Considering the ions involved in forming DCO$^+$ and DCN, it seems reasonable that their formation will be enhanced in regions that are irradiated, but not heated by FUV photons or X-rays. To test this hypothesis requires  H$^{13}$CO$^+$ abundances toward both systems (to measure whether the DCO$^+$/HCO$^+$ abundance ratio varies as well) in combination with a model that simultaneously treat deuterium chemistry and UV and X-ray radiative transfer.

H$_2$CO can form both through gas and grain surface processes. The gas phase process starts with CH$_3^+$ reacting with H$_2$ \citep{Roberts07b} and is expected to be at least as efficient in disks around low and intermediate mass stars. In contrast, H$_2$CO formation on grains requires the freeze-out of CO, which is only efficient at low temperatures. The absence of H$_2$CO toward the more luminous stars suggests that the grain surface formation mechanism dominates in disks and it is also another indication 
of the lack of  a large cold chemistry reservoir toward disks around intermediate mass stars.
It also suggests that the organic molecules formed in the protostellar stage, 
where H$_2$CO is common, do not survive in the gas phase in mature disks. 

In summary, all potential tracers of cold chemistry imply the same lack of cold disk material around Herbig Ae stars, which is in agreement with a recent survey of CO gas toward Herbig Ae/Be stars \citep{Panic09}. In contrast, \citet{Pietu07} find that the disk around the Herbig Ae star MWC 480 contains large amounts of cold CO gas, below 17~K, indicative of cold material outside of 200 AU. At these temperatures CO should not be in the gas phase at all, however, since it is below the sublimation point of CO ice. Its presence is a sign of either efficient mixing in the disk or efficient non-thermal ice evaporation, perhaps through photodesorption \citep{Oberg07b,Hersant09}. Mixing may drag up material from the midplane on shorter timescales than the cold chemistry timescales, explaining the lack of cold chemistry tracers. Efficient photodesorption of CO into the gas phase would also explain the lack of N$_2$H$^+$ and H$_2$CO, while its impact on the deuterium fractionation is harder to assess. The same processes are probably present in disks around T Tauri stars as well, but because their disks are overall colder there is still enough material protected from vertical mixing and photodesorption on long enough timescales for large amounts of N$_2$H$^+$, DCO$^+$ and H$_2$CO to form.

The lack of CH$_3$OH detections does not put strong 
constraints on the CH$_3$OH/H$_2$CO abundance ratio, since H$_2$CO is barely 
detected and the CH$_3$OH transitions in this spectral region are more than an 
order of magnitude weaker than the observed H$_2$CO transitions. To put stronger constraints on CH$_3$OH abundances in disks instead requires targeted observations of the most intense CH$_3$OH lines.

\section{Conclusions}

Protoplanetary disks exhibit a rich chemistry that varies significantly between different objects within the same star forming region. Some of this variation can be understood in terms of the central star and its heating of the disk -- the cold chemistry tracers N$_2$H$^+$, DCO$^+$, DCN and H$_2$CO are only detected toward T Tauri stars in our disk sample of four T Tauri stars and two Herbig Ae stars. Tracers of photochemistry, especially CN and HCN, show no clear dependence on quiescent stellar luminosity within the sample. Deuterium fractionation also seems to depend on parameters other than the disk temperature structure. For these chemical systems, the impact of other sources of irradiation, e.g. accretion shocks and X-rays, as well as the disk structure and grain characteristics may all be more important for the chemical evolution than the quiescent stellar luminosity. Investigating the relative importance of these different disk and star characteristics requires a combination of detailed modeling of the current sample, an increase in the number of sources to boost the statistics and span more parameters -- especially a larger range of accretion rates and disks around the intermediate F stars -- and targeted observations of rare isotopes of CO and HCO$^+$ to extract accurate abundance ratios. While the chemical evolution in protoplanetary disks is clearly complex, the qualitative agreement between at least parts of the early DISCS results and our current chemical understanding is promising for the ongoing modeling of these objects. The key results so far are listed below.

\begin{enumerate}
\item Six disks in Taurus (DM Tau, AA Tau, LkCa 15, GM Aur, CQ Tau and MWC 480) have been surveyed for 10 molecules, CO, HCO$^+$, DCO$^+$, CN, HCN, DCN, H$_2$CO, N$_2$H$^+$, CH$_3$OH and c-C$_3$H$_2$, with a high detection rate and large chemical variability.
\item The brightest molecular lines, CO 2-1, HCO$^+$ 3-2, CN 3-2  and HCN 3-2 are detected toward all disks, except for HCN toward CQ Tau. Other molecular lines tracing different types of cold chemistry, N$_2$H$^+$, DCO$^+$, DCN and H$_2$CO, are only detected toward disks around T Tauri stars, indicative of a lack of cold regions around Herbig Ae stars for long enough time scales.
\item Both the absolute CN flux and the CN/HCN ratio vary significantly among the observed disks and their variation seems independent of stellar luminosity, suggestive of that other parameters such as accretion luminosity and dust growth and dust settling play an important role for the chemical evolution in disks.
\item Among the cold chemistry tracers the DCO$^+$/N$_2$H$^+$ ratio varies by an order of magnitude suggesting that the deuterium fractionation depends on other parameters, including the radiation field, beyond the amount of cold material present in the disk.
\end{enumerate}

{\it Facilities:} \facility{SMA}

\acknowledgments

This work has benefitted from discussions with and comments from Ewine van Dishoeck, Geoffrey Blake and Michiel Hogerheijde, and from a helpful review by  an anonymous referee. The SMA is a joint project between the Smithsonian Astrophysical Observatory and the Academia Sinica Institute of Astronomy and Astrophysics and is funded by the Smithsonian Institution and the Academia Sinica. Support for K.~I.~O. and S.~M.~A. is provided by NASA through Hubble Fellowship grants  awarded by the Space Telescope Science Institute, which is operated by the Association of Universities for Research in Astronomy, Inc., for NASA, under contract NAS 5-26555.  C.~E. was supported by the National Science Foundation under Award No.~0901947. E.~A.~B. acknowledges support by NSF Grant \#0707777

%\bibliographystyle{aa}
%\bibliography{mybib}

\begin{thebibliography}{80}

\bibitem[{{Acke} {et~al.}(2004){Acke}, {van den Ancker}, {Dullemond}, {van
  Boekel}, \& {Waters}}]{Acke04}
{Acke}, B., {van den Ancker}, M.~E., {Dullemond}, C.~P., {van Boekel}, R., \&
  {Waters}, L.~B.~F.~M. 2004, \aap, 422, 621

\bibitem[{{Adams} {et~al.}(1990){Adams}, {Emerson}, \& {Fuller}}]{Adams90}
{Adams}, F.~C., {Emerson}, J.~P., \& {Fuller}, G.~A. 1990, \apj, 357, 606

\bibitem[{{Aikawa} \& {Herbst}(1999)}]{Aikawa99}
{Aikawa}, Y. \& {Herbst}, E. 1999, \aap, 351, 233

\bibitem[{{Alencar} \& {Batalha}(2002)}]{Alencar02}
{Alencar}, S.~H.~P. \& {Batalha}, C. 2002, \apj, 571, 378

\bibitem[{{Andrews} \& {Williams}(2005)}]{Andrews05}
{Andrews}, S.~M. \& {Williams}, J.~P. 2005, \apj, 631, 1134

\bibitem[{{Andrews} \& {Williams}(2007)}]{Andrews07}
{Andrews}, S.~M. \& {Williams}, J.~P. 2007, \apj, 659, 705

\bibitem[{{Beckwith} \& {Sargent}(1991)}]{Beckwith91}
{Beckwith}, S.~V.~W. \& {Sargent}, A.~I. 1991, \apj, 381, 250

\bibitem[{{Beckwith} {et~al.}(1990){Beckwith}, {Sargent}, {Chini}, \&
  {Guesten}}]{Beckwith90}
{Beckwith}, S.~V.~W., {Sargent}, A.~I., {Chini}, R.~S., \& {Guesten}, R. 1990,
  \aj, 99, 924

\bibitem[{{Belloche} {et~al.}(2009){Belloche}, {Garrod}, {M{\"u}ller},
  {Menten}, {Comito}, \& {Schilke}}]{Belloche09}
{Belloche}, A., {Garrod}, R.~T., {M{\"u}ller}, H.~S.~P., {et~al.} 2009, \aap,
  499, 215

\bibitem[{{Bergin} {et~al.}(2003){Bergin}, {Calvet}, {D'Alessio}, \&
  {Herczeg}}]{Bergin03}
{Bergin}, E., {Calvet}, N., {D'Alessio}, P., \& {Herczeg}, G.~J. 2003, \apjl,
  591, L159

\bibitem[{{Bergin} {et~al.}(2004){Bergin}, {Calvet}, {Sitko}, {Abgrall},
  {D'Alessio}, {Herczeg}, {Roueff}, {Qi}, {Lynch}, {Russell}, {Brafford}, \&
  {Perry}}]{Bergin04}
{Bergin}, E., {Calvet}, N., {Sitko}, M.~L., {et~al.} 2004, \apjl, 614, L133

\bibitem[{{Bergin} {et~al.}(2002){Bergin}, {Alves}, {Huard}, \&
  {Lada}}]{Bergin02}
{Bergin}, E.~A., {Alves}, J., {Huard}, T., \& {Lada}, C.~J. 2002, \apjl, 570,
  L101

\bibitem[{{Calvet} {et~al.}(2005){Calvet}, {D'Alessio}, {Watson},
  {Franco-Hern{\'a}ndez}, {Furlan}, {Green}, {Sutter}, {Forrest}, {Hartmann},
  {Uchida}, {Keller}, {Sargent}, {Najita}, {Herter}, {Barry}, \&
  {Hall}}]{Calvet05}
{Calvet}, N., {D'Alessio}, P., {Watson}, D.~M., {et~al.} 2005, \apjl, 630, L185

\bibitem[{{Calvet} {et~al.}(2004){Calvet}, {Muzerolle}, {Brice{\~n}o},
  {Hern{\'a}ndez}, {Hartmann}, {Saucedo}, \& {Gordon}}]{Calvet04}
{Calvet}, N., {Muzerolle}, J., {Brice{\~n}o}, C., {et~al.} 2004, \aj, 128, 1294

\bibitem[{{Chapillon} {et~al.}(2008){Chapillon}, {Guilloteau}, {Dutrey}, \&
  {Pi{\'e}tu}}]{Chapillon08}
{Chapillon}, E., {Guilloteau}, S., {Dutrey}, A., \& {Pi{\'e}tu}, V. 2008, \aap,
  488, 565

\bibitem[{{Chiang} {et~al.}(2001){Chiang}, {Joung}, {Creech-Eakman}, {Qi},
  {Kessler}, {Blake}, \& {van Dishoeck}}]{Chiang01}
{Chiang}, E.~I., {Joung}, M.~K., {Creech-Eakman}, M.~J., {et~al.} 2001, \apj,
  547, 1077

\bibitem[{{Crovisier} {et~al.}(2004){Crovisier}, {Bockel{\'e}e-Morvan},
  {Biver}, {Colom}, {Despois}, \& {Lis}}]{Crovisier04}
{Crovisier}, J., {Bockel{\'e}e-Morvan}, D., {Biver}, N., {et~al.} 2004, \aap,
  418, L35

\bibitem[{{Dartois} {et~al.}(2003){Dartois}, {Dutrey}, \&
  {Guilloteau}}]{Dartois03}
{Dartois}, E., {Dutrey}, A., \& {Guilloteau}, S. 2003, \aap, 399, 773

\bibitem[{{Dutrey}(2004)}]{Dutrey04}
{Dutrey}, A. 2004, in IAU Symposium, Vol. 221, Star Formation at High Angular
  Resolution, ed. {M.~G.~Burton, R.~Jayawardhana, \& T.~L.~Bourke}, 411--+

\bibitem[{{Dutrey} {et~al.}(1996){Dutrey}, {Guilloteau}, {Duvert}, {Prato},
  {Simon}, {Schuster}, \& {Menard}}]{Dutrey96}
{Dutrey}, A., {Guilloteau}, S., {Duvert}, G., {et~al.} 1996, \aap, 309, 493

\bibitem[{{Dutrey} {et~al.}(1997){Dutrey}, {Guilloteau}, \&
  {Guelin}}]{Dutrey97}
{Dutrey}, A., {Guilloteau}, S., \& {Guelin}, M. 1997, \aap, 317, L55

\bibitem[{{Dutrey} {et~al.}(2008){Dutrey}, {Guilloteau}, {Pi{\'e}tu},
  {Chapillon}, {Gueth}, {Henning}, {Launhardt}, {Pavlyuchenkov}, {Schreyer}, \&
  {Semenov}}]{Dutrey08}
{Dutrey}, A., {Guilloteau}, S., {Pi{\'e}tu}, V., {et~al.} 2008, \aap, 490, L15

\bibitem[{{Dutrey} {et~al.}(1998){Dutrey}, {Guilloteau}, {Prato}, {Simon},
  {Duvert}, {Schuster}, \& {Menard}}]{Dutrey98}
{Dutrey}, A., {Guilloteau}, S., {Prato}, L., {et~al.} 1998, \aap, 338, L63

\bibitem[{{Dutrey} {et~al.}(2007){Dutrey}, {Henning}, {Guilloteau}, {Semenov},
  {Pi{\'e}tu}, {Schreyer}, {Bacmann}, {Launhardt}, {Pety}, \&
  {Gueth}}]{Dutrey07}
{Dutrey}, A., {Henning}, T., {Guilloteau}, S., {et~al.} 2007, \aap, 464, 615

\bibitem[{{Duvert} {et~al.}(2000){Duvert}, {Guilloteau}, {M{\'e}nard}, {Simon},
  \& {Dutrey}}]{Duvert00}
{Duvert}, G., {Guilloteau}, S., {M{\'e}nard}, F., {Simon}, M., \& {Dutrey}, A.
  2000, \aap, 355, 165

\bibitem[{{Espaillat} {et~al.}(2007){Espaillat}, {Calvet}, {D'Alessio},
  {Hern{\'a}ndez}, {Qi}, {Hartmann}, {Furlan}, \& {Watson}}]{Espaillat07}
{Espaillat}, C., {Calvet}, N., {D'Alessio}, P., {et~al.} 2007, \apjl, 670, L135

\bibitem[{{Furlan} {et~al.}(2009){Furlan}, {Watson}, {McClure}, {Manoj},
  {Espaillat}, {D'Alessio}, {Calvet}, {Kim}, {Sargent}, {Forrest}, \&
  {Hartmann}}]{Furlan09}
{Furlan}, E., {Watson}, D.~M., {McClure}, M.~K., {et~al.} 2009, \apj, 703, 1964

\bibitem[{{Garcia Lopez} {et~al.}(2006){Garcia Lopez}, {Natta}, {Testi}, \&
  {Habart}}]{GarciaLopez06}
{Garcia Lopez}, R., {Natta}, A., {Testi}, L., \& {Habart}, E. 2006, \aap, 459,
  837

\bibitem[{{Glassgold} {et~al.}(2005){Glassgold}, {Feigelson}, {Montmerle}, \&
  {Wolk}}]{Glassgold05}
{Glassgold}, A.~E., {Feigelson}, E.~D., {Montmerle}, T., \& {Wolk}, S. 2005, in
  Astronomical Society of the Pacific Conference Series, Vol. 341, Chondrites
  and the Protoplanetary Disk, ed. {A.~N.~Krot, E.~R.~D.~Scott, \&
  B.~Reipurth}, 165--+

\bibitem[{{Glassgold} {et~al.}(1997){Glassgold}, {Najita}, \&
  {Igea}}]{Glassgold97}
{Glassgold}, A.~E., {Najita}, J., \& {Igea}, J. 1997, \apj, 480, 344

\bibitem[{{G{\"u}del} \& {Naz{\'e}}(2009)}]{Gudel09}
{G{\"u}del}, M. \& {Naz{\'e}}, Y. 2009, \aapr, 17, 309

\bibitem[{{Guilloteau} \& {Dutrey}(1998)}]{Guilloteau98}
{Guilloteau}, S. \& {Dutrey}, A. 1998, \aap, 339, 467

\bibitem[{{Guilloteau} {et~al.}(2006){Guilloteau}, {Pi{\'e}tu}, {Dutrey}, \&
  {Gu{\'e}lin}}]{Guilloteau06}
{Guilloteau}, S., {Pi{\'e}tu}, V., {Dutrey}, A., \& {Gu{\'e}lin}, M. 2006,
  \aap, 448, L5

\bibitem[{{Hamidouche} {et~al.}(2006){Hamidouche}, {Looney}, \&
  {Mundy}}]{Hamidouche06}
{Hamidouche}, M., {Looney}, L.~W., \& {Mundy}, L.~G. 2006, \apj, 651, 321

\bibitem[{{Henning} {et~al.}(2010){Henning}, {Semenov}, {Guilloteau}, {Dutrey},
  {Hersant}, {Wakelam}, {Chapillon}, {Launhardt}, {Pietu}, \&
  {Schreyer}}]{Henning10}
{Henning}, T., {Semenov}, D., {Guilloteau}, S., {et~al.} 2010, ArXiv e-prints

\bibitem[{{Herczeg} {et~al.}(2002){Herczeg}, {Linsky}, {Valenti},
  {Johns-Krull}, \& {Wood}}]{Herczeg02}
{Herczeg}, G.~J., {Linsky}, J.~L., {Valenti}, J.~A., {Johns-Krull}, C.~M., \&
  {Wood}, B.~E. 2002, \apj, 572, 310

\bibitem[{{Hersant} {et~al.}(2009){Hersant}, {Wakelam}, {Dutrey}, {Guilloteau},
  \& {Herbst}}]{Hersant09}
{Hersant}, F., {Wakelam}, V., {Dutrey}, A., {Guilloteau}, S., \& {Herbst}, E.
  2009, \aap, 493, L49

\bibitem[{{Hughes} {et~al.}(2009){Hughes}, {Andrews}, {Espaillat}, {Wilner},
  {Calvet}, {D'Alessio}, {Qi}, {Williams}, \& {Hogerheijde}}]{Hughes09}
{Hughes}, A.~M., {Andrews}, S.~M., {Espaillat}, C., {et~al.} 2009, \apj, 698,
  131

\bibitem[{{Isella} {et~al.}(2009){Isella}, {Carpenter}, \&
  {Sargent}}]{Isella09}
{Isella}, A., {Carpenter}, J.~M., \& {Sargent}, A.~I. 2009, \apj, 701, 260

\bibitem[{{Jonkheid} {et~al.}(2007){Jonkheid}, {Dullemond}, {Hogerheijde}, \&
  {van Dishoeck}}]{Jonkheid07}
{Jonkheid}, B., {Dullemond}, C.~P., {Hogerheijde}, M.~R., \& {van Dishoeck},
  E.~F. 2007, \aap, 463, 203

\bibitem[{{Kastner} {et~al.}(2008){Kastner}, {Zuckerman}, {Hily-Blant}, \&
  {Forveille}}]{Kastner08}
{Kastner}, J.~H., {Zuckerman}, B., {Hily-Blant}, P., \& {Forveille}, T. 2008,
  \aap, 492, 469

\bibitem[{{Kessler-Silacci}(2004)}]{KesslerSilacci04}
{Kessler-Silacci}, J. 2004, PhD thesis, California Institute of Technology

\bibitem[{{Kitamura} {et~al.}(2002){Kitamura}, {Momose}, {Yokogawa}, {Kawabe},
  {Tamura}, \& {Ida}}]{Kitamura02}
{Kitamura}, Y., {Momose}, M., {Yokogawa}, S., {et~al.} 2002, \apj, 581, 357

\bibitem[{{Koerner} {et~al.}(1993){Koerner}, {Sargent}, \&
  {Beckwith}}]{Koerner93}
{Koerner}, D.~W., {Sargent}, A.~I., \& {Beckwith}, S.~V.~W. 1993, Icarus, 106,
  2

\bibitem[{{Looney} {et~al.}(2000){Looney}, {Mundy}, \& {Welch}}]{Looney00}
{Looney}, L.~W., {Mundy}, L.~G., \& {Welch}, W.~J. 2000, \apj, 529, 477

\bibitem[{{Mannings}(1994)}]{Mannings94}
{Mannings}, V. 1994, \mnras, 271, 587

\bibitem[{{Mannings} \& {Sargent}(1997)}]{Mannings97}
{Mannings}, V. \& {Sargent}, A.~I. 1997, \apj, 490, 792

\bibitem[{{Mannings} \& {Sargent}(2000)}]{Mannings00}
{Mannings}, V. \& {Sargent}, A.~I. 2000, \apj, 529, 391

\bibitem[{{Markwick} {et~al.}(2002){Markwick}, {Ilgner}, {Millar}, \&
  {Henning}}]{Markwick02}
{Markwick}, A.~J., {Ilgner}, M., {Millar}, T.~J., \& {Henning}, T. 2002, \aap,
  385, 632

\bibitem[{{M{\'e}nard} {et~al.}(2003){M{\'e}nard}, {Bouvier}, {Dougados},
  {Mel'nikov}, \& {Grankin}}]{Menard03}
{M{\'e}nard}, F., {Bouvier}, J., {Dougados}, C., {Mel'nikov}, S.~Y., \&
  {Grankin}, K.~N. 2003, \aap, 409, 163

\bibitem[{{M{\"u}ller} {et~al.}(2001){M{\"u}ller}, {Thorwirth}, {Roth}, \&
  {Winnewisser}}]{Muller01}
{M{\"u}ller}, H.~S.~P., {Thorwirth}, S., {Roth}, D.~A., \& {Winnewisser}, G.
  2001, \aap, 370, L49

\bibitem[{{Natta} {et~al.}(2001){Natta}, {Prusti}, {Neri}, {Wooden}, {Grinin},
  \& {Mannings}}]{Natta01}
{Natta}, A., {Prusti}, T., {Neri}, R., {et~al.} 2001, \aap, 371, 186

\bibitem[{{{\"O}berg} {et~al.}(2007){{\"O}berg}, {Fuchs}, {Awad}, {Fraser},
  {Schlemmer}, {van Dishoeck}, \& {Linnartz}}]{Oberg07b}
{{\"O}berg}, K.~I., {Fuchs}, G.~W., {Awad}, Z., {et~al.} 2007, \apjl, 662, L23

\bibitem[{{{\"O}berg} {et~al.}(2009){{\"O}berg}, {Garrod}, {van Dishoeck}, \&
  {Linnartz}}]{Oberg09d}
{{\"O}berg}, K.~I., {Garrod}, R.~T., {van Dishoeck}, E.~F., \& {Linnartz}, H.
  2009, \aap, 504, 891

\bibitem[{{{\"O}berg} {et~al.}(2005){{\"O}berg}, {van Broekhuizen}, {Fraser},
  {Bisschop}, {van Dishoeck}, \& {Schlemmer}}]{Oberg05}
{{\"O}berg}, K.~I., {van Broekhuizen}, F., {Fraser}, H.~J., {et~al.} 2005,
  \apjl, 621, L33

\bibitem[{{Osterloh} \& {Beckwith}(1995)}]{Osterloh95}
{Osterloh}, M. \& {Beckwith}, S.~V.~W. 1995, \apj, 439, 288

\bibitem[{{Pani{\'c}} \& {Hogerheijde}(2009)}]{Panic09}
{Pani{\'c}}, O. \& {Hogerheijde}, M.~R. 2009, \aap, 508, 707

\bibitem[{{Pascucci} {et~al.}(2009){Pascucci}, {Apai}, {Luhman}, {Henning},
  {Bouwman}, {Meyer}, {Lahuis}, \& {Natta}}]{Pascucci09}
{Pascucci}, I., {Apai}, D., {Luhman}, K., {et~al.} 2009, \apj, 696, 143

\bibitem[{{Pi{\'e}tu} {et~al.}(2007){Pi{\'e}tu}, {Dutrey}, \&
  {Guilloteau}}]{Pietu07}
{Pi{\'e}tu}, V., {Dutrey}, A., \& {Guilloteau}, S. 2007, \aap, 467, 163

\bibitem[{{Pi{\'e}tu} {et~al.}(2006){Pi{\'e}tu}, {Dutrey}, {Guilloteau},
  {Chapillon}, \& {Pety}}]{Pietu06}
{Pi{\'e}tu}, V., {Dutrey}, A., {Guilloteau}, S., {Chapillon}, E., \& {Pety}, J.
  2006, \aap, 460, L43

\bibitem[{{Qi} {et~al.}(2003){Qi}, {Kessler}, {Koerner}, {Sargent}, \&
  {Blake}}]{Qi03}
{Qi}, C., {Kessler}, J.~E., {Koerner}, D.~W., {Sargent}, A.~I., \& {Blake},
  G.~A. 2003, \apj, 597, 986

\bibitem[{{Qi} {et~al.}(2008){Qi}, {Wilner}, {Aikawa}, {Blake}, \&
  {Hogerheijde}}]{Qi08}
{Qi}, C., {Wilner}, D.~J., {Aikawa}, Y., {Blake}, G.~A., \& {Hogerheijde},
  M.~R. 2008, \apj, 681, 1396

\bibitem[{{Ricci} {et~al.}(2009){Ricci}, {Testi}, {Natta}, {Neri}, {Cabrit}, \&
  {Herczeg}}]{Ricci09}
{Ricci}, L., {Testi}, L., {Natta}, A., {et~al.} 2009, ArXiv e-prints

\bibitem[{{Roberts} \& {Millar}(2000)}]{Roberts00}
{Roberts}, H. \& {Millar}, T.~J. 2000, \aap, 361, 388

\bibitem[{{Roberts} \& {Millar}(2007)}]{Roberts07b}
{Roberts}, H. \& {Millar}, T.~J. 2007, \aap, 471, 849

\bibitem[{{Rodmann} {et~al.}(2006){Rodmann}, {Henning}, {Chandler}, {Mundy}, \&
  {Wilner}}]{Rodmann06}
{Rodmann}, J., {Henning}, T., {Chandler}, C.~J., {Mundy}, L.~G., \& {Wilner},
  D.~J. 2006, \aap, 446, 211

\bibitem[{{Salyk} {et~al.}(2009){Salyk}, {Blake}, {Boogert}, \&
  {Brown}}]{Salyk09}
{Salyk}, C., {Blake}, G.~A., {Boogert}, A.~C.~A., \& {Brown}, J.~M. 2009, \apj,
  699, 330

\bibitem[{{Schreyer} {et~al.}(2008){Schreyer}, {Guilloteau}, {Semenov},
  {Bacmann}, {Chapillon}, {Dutrey}, {Gueth}, {Henning}, {Hersant}, {Launhardt},
  {Pety}, \& {Pi{\'e}tu}}]{Schreyer08}
{Schreyer}, K., {Guilloteau}, S., {Semenov}, D., {et~al.} 2008, \aap, 491, 821

\bibitem[{{Simon} {et~al.}(2000){Simon}, {Dutrey}, \& {Guilloteau}}]{Simon00}
{Simon}, M., {Dutrey}, A., \& {Guilloteau}, S. 2000, \apj, 545, 1034

\bibitem[{{Testi} {et~al.}(2001){Testi}, {Natta}, {Shepherd}, \&
  {Wilner}}]{Testi01}
{Testi}, L., {Natta}, A., {Shepherd}, D.~S., \& {Wilner}, D.~J. 2001, \apj,
  554, 1087

\bibitem[{{Testi} {et~al.}(2003){Testi}, {Natta}, {Shepherd}, \&
  {Wilner}}]{Testi03}
{Testi}, L., {Natta}, A., {Shepherd}, D.~S., \& {Wilner}, D.~J. 2003, \aap,
  403, 323

\bibitem[{{The} {et~al.}(1994){The}, {de Winter}, \& {Perez}}]{The94}
{The}, P.~S., {de Winter}, D., \& {Perez}, M.~R. 1994, \aaps, 104, 315

\bibitem[{{Thi} {et~al.}(2004){Thi}, {van Zadelhoff}, \& {van
  Dishoeck}}]{Thi04}
{Thi}, W., {van Zadelhoff}, G., \& {van Dishoeck}, E.~F. 2004, \aap, 425, 955

\bibitem[{{van Dishoeck} {et~al.}(1995){van Dishoeck}, {Blake}, {Jansen}, \&
  {Groesbeck}}]{vanDishoeck95}
{van Dishoeck}, E.~F., {Blake}, G.~A., {Jansen}, D.~J., \& {Groesbeck}, T.~D.
  1995, \apj, 447, 760

\bibitem[{{van Dishoeck} {et~al.}(2003){van Dishoeck}, {Thi}, \& {van
  Zadelhoff}}]{vanDishoeck03}
{van Dishoeck}, E.~F., {Thi}, W., \& {van Zadelhoff}, G. 2003, \aap, 400, L1

\bibitem[{{van Zadelhoff} {et~al.}(2003){van Zadelhoff}, {Aikawa},
  {Hogerheijde}, \& {van Dishoeck}}]{vanZadelhoff03}
{van Zadelhoff}, G., {Aikawa}, Y., {Hogerheijde}, M.~R., \& {van Dishoeck},
  E.~F. 2003, \aap, 397, 789

\bibitem[{{van Zadelhoff} {et~al.}(2001){van Zadelhoff}, {van Dishoeck}, {Thi},
  \& {Blake}}]{vanZadelhoff01}
{van Zadelhoff}, G., {van Dishoeck}, E.~F., {Thi}, W., \& {Blake}, G.~A. 2001,
  \aap, 377, 566

\bibitem[{{Weintraub} {et~al.}(1989){Weintraub}, {Sandell}, \&
  {Duncan}}]{Weintraub89}
{Weintraub}, D.~A., {Sandell}, G., \& {Duncan}, W.~D. 1989, \apjl, 340, L69

\bibitem[{{White} \& {Ghez}(2001)}]{White01}
{White}, R.~J. \& {Ghez}, A.~M. 2001, \apj, 556, 265

\bibitem[{{Willacy}(2007)}]{Willacy07}
{Willacy}, K. 2007, \apj, 660, 441

\end{thebibliography}

\begin{deluxetable}{lcccccc}
\tabletypesize{\scriptsize}
\tablecaption{Central star and accretion data. \label{tbl:star}}
\tablewidth{0pt}
\tablehead{
\colhead{Source} & \colhead{RA} & \colhead{DEC} & \colhead{Sp. type}& \colhead{L$_*$ (L$_{\odot}$)} & \colhead{$\dot{M}$ (10$^{-9}$ M$_{\odot}$ yr$^{-1}$)}}
\startdata
DM Tau     &04 33 48.73  & $+$18 10 10.0& M1.5\tablenotemark{a}  & 0.3\tablenotemark{a} &0.8\tablenotemark{b}\\

AA Tau & 04 34 55.42  & $+$24 28 53.2& K7\tablenotemark{c}  &0.59\tablenotemark{c} & 7\tablenotemark{c} \\

LkCa 15   	& 04 39 17.78  & $+$22 21 03.5 & K3\tablenotemark{a}  &1.2\tablenotemark{a} &8\tablenotemark{b}  \\

GM Aur    & 04 55 10.98  & $+$30 21 59.4 & K5\tablenotemark{a} &0.93 \tablenotemark{a}   &3\tablenotemark{b}\\

CQ Tau  &  05 35 58.47 & $+$24 44 54.1& F2\tablenotemark{d} & 8.0\tablenotemark{e}  & $<$8\tablenotemark{f}  \\

MWC 480  & 04 58 46.27  & $+$29 50 37.0 & A3\tablenotemark{g} &11.5\tablenotemark{g}  & \nodata \\
\enddata
\tablenotetext{a}{Espaillat et al. (submitted)}
\tablenotetext{b}{\citet{Bergin04}}
\tablenotetext{c}{\citet{White01}}
\tablenotetext{d}{\citet{The94}}
\tablenotetext{e}{\citet{Mannings97}}
\tablenotetext{f}{\citet{GarciaLopez06}}
\tablenotetext{g}{\citet{Simon00}}
\end{deluxetable}

\begin{deluxetable}{lcccccc}
\tabletypesize{\scriptsize}
\tablecaption{Disk data. \label{tbl:disk}}
\tablewidth{0pt}
\tablehead{
\colhead{Source} & \colhead{PA$_{\rm CO}$ (deg)} & \colhead{$i_{\rm CO}$ (deg)} &\colhead{R$_{\rm CO}$ (AU) }& \colhead{R$_{\rm hole}$ (AU) }& \colhead{M$_{\rm D}$ (M$_{\odot}$)} & \colhead{$\beta_{mm}$}}
\startdata
DM Tau & -25\tablenotemark{a}& -35\tablenotemark{a}& 890\tablenotemark{a}&3\tablenotemark{c} & 0.02\tablenotemark{d}&1.0\tablenotemark{e}, 1.5\tablenotemark{f}\\

AA Tau & -89\tablenotemark{g} & 75\tablenotemark{h} &995\tablenotemark{g}  & \nodata & 0.01\tablenotemark{d}  & 0.3\tablenotemark{e}, 1.0\tablenotemark{f} \\

GM Aur  &46\tablenotemark{b}   & 49\tablenotemark{b} &445\tablenotemark{b} &24\tablenotemark{c} & 0.03\tablenotemark{d} &1.0\tablenotemark{e}, 1.2\tablenotemark{f} \\

LkCa 15 & 61\tablenotemark{a}   &52\tablenotemark{a} &905\tablenotemark{a}&58\tablenotemark{i} &0.05\tablenotemark{d}&\nodata \\

CQ Tau& -43\tablenotemark{j}  & 29\tablenotemark{j}  & 200\tablenotemark{j} &\nodata&0.009\tablenotemark{k}  &0.7\tablenotemark{j}, 0.6\tablenotemark{l}\\

MWC 480 &-32\tablenotemark{a}  &38\tablenotemark{a} &740\tablenotemark{a} &\nodata &0.11\tablenotemark{m}  &0.8\tablenotemark{n}, 1.4\tablenotemark{l}\\
\enddata
\tablenotetext{a}{\citet{Pietu07} -- the reported PAs have been converted to conform with its conventional definition as the angle of the projected disk major axis.}
\tablenotetext{b}{\citet{Dutrey08}}
\tablenotetext{c}{\citet{Calvet05}}
\tablenotetext{d}{\citet{Andrews05}}
\tablenotetext{e}{\citet{Ricci09}}
\tablenotetext{f}{\citet{Andrews07}}
\tablenotetext{g}{\citet{KesslerSilacci04}}
\tablenotetext{h}{\citet{KesslerSilacci04}}
\tablenotetext{i}{Espaillat et al. (submitted)}
\tablenotetext{j}{\citet{Chapillon08} -- the reported PA has been converted to conform with its conventional definition as the angle of the projected disk major axis.}
\tablenotetext{k}{\citet{Mannings00}}
\tablenotetext{l}{\citet{Testi03}}
\tablenotetext{m}{\citet{Chiang01}}
\tablenotetext{n}{\citet{Hamidouche06}}
\end{deluxetable}

\begin{deluxetable}{lcccccc}
\tabletypesize{\scriptsize}
\tablecaption{\small Spectral setup: 1.4 mm setting (214.670--218.653 GHz,
LSB and 226.671--230.654 GHz, USB). \label{tbl:setup1}}
\tablewidth{0pt}
\tablehead{
\colhead{Chunk } & \colhead{Frequency range} & \colhead{Channels} & \colhead{Resolution}& \colhead{Lines} & \colhead{Frequency}& \colhead{T$_{\rm up}$}\\
 & (GHz) & & (km s$^{-1}$)& &(GHz)&(K)
}
\startdata
  &                 &      &  LSB              &    &        &  \\
\tableline
S31 & 216.064--216.168 & 256 & 0.56 & DCO$^+$ 3--2 & 216.1126 & 21 \\
S18 & 217.155--217.259 & 256 & 0.56 & DCN 3--2     & 217.2386 & 21 \\
S10 & 217.811--217.915 & 256 & 0.56 & $c$-C$_3$H$_2$ $6_{1\:6}-5_{0\:5}$  & 217.8221 & 39 \\
S06 & 218.140--218.244 & 64  & 2.23 & H$_2$CO $3_{0\:3}-2_{0\:2}$ &
218.2222 & 21\\
S03 & 218.392--218.496 & 256 & 0.56 & CH$_3$OH-E $4_{2\:1}-3_{1\:2}$ &
218.4401 & 46 \\
\tableline
 &                 &      &  USB                &   &        &  \\
\tableline
S01 & 226.671--226.775 & 128 & 1.07 & CN $2_{2\:2}-1_{1\:1}$ &
226.6794 & 16 \\
S03 & 226.828--226.932 & 256 & 0.54 & CN $2_{3\:3/4/2}-1_{2\:2/3/1}$ &
226.8747 & 16 \\
S47 & 230.468--230.572 & 256\tablenotemark{a} & 0.53 & CO 2--1 & 230.538 & 17 \\
\enddata
\tablenotetext{a}{64 toward LkCa 15}
\end{deluxetable}

\begin{deluxetable}{lcccccc}
\tabletypesize{\scriptsize}
\tablecaption{\small Spectral setup: 1.1 mm setting (265.777--269.760 GHz,
LSB and 277.778--281.761 GHz, USB). \label{tbl:setup2}}
\tablewidth{0pt}
\tablehead{
\colhead{Chunk } & \colhead{Frequency range} & \colhead{Channels} & \colhead{Resolution}& \colhead{Lines} & \colhead{Frequency}& \colhead{T$_{\rm up}$}\\
 & (GHz) & & (km s$^{-1}$)& &(GHz)&(K)
}
\startdata
  &                 &      &  LSB              &    &        &  \\
\tableline
S47 & 265.858--265.962 & 256 & 0.46 & HCN 3--2 & 265.8862 & 26 \\ 
S27 & 267.499--267.603 & 256 & 0.46 & HCO$^+$ 3--2 & 267.5575 & 26 \\
\tableline
 &                 &      &  USB                &   &        &  \\
\tableline
S22 & 279.499--279.603 & 256 & 0.44 & N$_2$H$^+$ 3--2& 279.5117 & 27\\
S46 & 281.500--281.604 & 128\tablenotemark{a} & 0.87 & H$_2$CO $4_{1\:4}-3_{1\:3}$ &
281.5269 & 46 \\
\enddata
\tablenotetext{a}{64 toward LkCa 15}
\end{deluxetable}

\begin{deluxetable}{rccccc}
\tabletypesize{\scriptsize}
\tablecaption{\small SMA Observation Log. \label{tbl:obs}}
\tablewidth{0pt}
\tablehead{
\colhead{Date} & \colhead{Sources} & \colhead{Setting\tablenotemark{a}} & \colhead{Baselines\tablenotemark{b} (Antennas)}& \colhead{$\tau_{\rm 225GHz}$} & \colhead{T$_{sys}$(K)\tablenotemark{c}}
}
\startdata
2009 Nov 20 & LkCa 15, MWC 480 & 1.1 mm & 8--64(7) & 0.04-0.08 & 107--528 \\
     Nov 22 & LkCa 15& 1.4 mm & 9--52(6) & 0.04--0.1 & 74--103\\
     Nov 28 & MWC 480 & 1.4 mm & 10--108(6) & 0.04--0.09& 107--528 \\
     Dec 03 & DM Tau & 1.4 mm & 9--52(7) & 0.05--0.1 & 87--125 \\
     Dec 08 & DM Tau, GM Aur & 1.1 mm & 8--64(7)& 0.05--0.09 & 116--316\\
     Dec 09 & GM Aur & 1.4 mm & 7--52(7) & 0.06--0.1 & 87--216 \\
     Dec 15 & AA Tau, CQ Tau & 1.1 mm & 9--64(6) & 0.06--0.09 &
     111--287\\
     Dec 19 & CQ Tau & 1.4 mm & 9--58(8)& 0.06--0.15& 79--195 \\
     Dec 23 & AA Tau & 1.4 mm & 9--52(7)&0.05--0.09& 73--150 \\
     2010 Mar 22 & LkCa 15 & 1.1 mm & 4-34(7) & 0.06-0.1 & 83-151\\
\enddata
\tablenotetext{a}{See Tables \ref{tbl:setup1} and \ref{tbl:setup2}}
\tablenotetext{b}{Projected baseline in meters}
\tablenotetext{c}{Double sideband (DSB) system temperature}
\end{deluxetable}

\begin{deluxetable}{rccc}
\tabletypesize{\scriptsize}
\tablecaption{\small Beam sizes and position angles for each source and setting. \label{tbl:beam}}
\tablewidth{0pt}
\tablehead{
\colhead{Sources} & \colhead{Setting} & \colhead{Beam size (arc sec)} & \colhead{Beam PA (degrees)}
}
\startdata
DM Tau & 1.1 mm & $3.3\times2.6$&19\\
DM Tau & 1.4 mm & $3.9\times3.3$&19\\
AA Tau & 1.1 mm & $2.9\times2.5$&27\\
AA Tau & 1.4 mm & $3.9\times3.2$&10\\
LkCa 15 & 1.1 mm & $3.3\times2.7$&26\\
LkCa 15 & 1.4 mm & $3.8\times3.2$&14\\
GM Aur & 1.1 mm & $3.2\times2.7$&26\\
GM Aur & 1.4 mm & $3.8\times3.2$&14\\
CQ Tau & 1.1 mm & $2.9\times2.5$&27\\
CQ Tau & 1.4 mm & $3.5\times3.0$&-70\\
MWC 480 & 1.1 mm & $3.2\times2.5$&21\\
MWC 480& 1.4 mm & $3.8\times2.9$&-8\\
\enddata
\end{deluxetable}

\newpage

\begin{deluxetable}{lllllllllllllllllllll}
%\rotate
\tabletypesize{\scriptsize}
\tablecaption{\small Integrated spectra for DM Tau. Spectra for all sources are available on-line. \label{tbl:spec}}
\tablewidth{0pt}
\setlength{\tabcolsep}{0.04in}
\tablehead{
\multicolumn{2}{c}{CO 2-1} & \multicolumn{2}{c}{HCO$^+$ 3-2} & \multicolumn{2}{c}{DCO$^+$ 3-2} & \multicolumn{2}{c}{N$_2$H$^+$ 3-2} & \multicolumn{2}{c}{H$_2$CO 4-3}& \multicolumn{2}{c}{H$_2$CO 3-2}& \multicolumn{2}{c}{HCN 3-2}& \multicolumn{2}{c}{DCN 3-2} & \multicolumn{2}{c}{CN 2$_3$-1$_2$}& \multicolumn{2}{c}{CN 2$_2$-1$_1$}        \\
Vel.& Flux    & Vel.& Flux    & Vel.& Flux    & Vel.& Flux    & Vel.& Flux    & Vel.& Flux    & Vel.& Flux    & Vel.& Flux    & Vel.& Flux    & Vel.& Flux    \\
km/s    & Jy      & km/s    & Jy      & km/s    & Jy      & km/s    & Jy      & km/s    & Jy      & km/s    & Jy      & km/s    & Jy      & km/s    & Jy      & km/s    & Jy      & km/s    & Jy      
}
\startdata

-8.50    &0.09     &-8.70    &0.03     &-8.90    &-0.06    &-8.70    &0.07     &-8.80    &-0.01    &-7.70    &-0.00    &-8.70    &-0.04    &-8.50    &-0.04    &-8.60    &-0.00    &-8.60    &-0.02     \\
-8.00    &-0.07    &-8.20    &0.20     &-7.80    &0.07     &-7.80    &-0.06    &-7.90    &0.03     &-5.50    &0.01     &-8.30    &-0.01    &-7.40    &-0.06    &-8.10    &0.04     &-7.50    &-0.12     \\
-7.50    &0.05     &-7.80    &0.09     &-6.70    &-0.02    &-7.00    &0.09     &-7.00    &0.01     &-3.20    &0.03     &-7.80    &-0.14    &-6.20    &0.02     &-7.50    &0.09     &-6.40    &-0.07     \\
-7.00    &0.19     &-7.30    &-0.63    &-5.50    &-0.03    &-6.10    &0.09     &-6.20    &-0.20    &-1.00    &0.03     &-7.40    &0.16     &-5.10    &0.03     &-7.00    &0.06     &-5.30    &-0.17     \\
-6.40    &-0.23    &-6.90    &0.01     &-4.40    &-0.02    &-5.20    &-0.31    &-5.30    &-0.10    &1.20     &0.00     &-6.90    &0.04     &-4.00    &-0.01    &-6.40    &-0.01    &-4.30    &0.01      \\
-5.90    &0.13     &-6.40    &-0.09    &-3.30    &0.08     &-4.30    &0.12     &-4.40    &0.23     &3.40     &0.03     &-6.40    &-0.01    &-2.90    &-0.04    &-5.90    &0.29     &-3.20    &-0.05     \\
-5.40    &-0.12    &-6.00    &0.22     &-2.20    &0.04     &-3.50    &-0.19    &-3.60    &0.07     &5.70     &0.10     &-6.00    &0.58     &-1.80    &-0.04    &-5.40    &-0.00    &-2.10    &-0.18     \\
-4.80    &-0.13    &-5.50    &0.35     &-1.00    &-0.02    &-2.60    &0.02     &-2.70    &0.06     &7.90     &0.02     &-5.50    &-0.55    &-0.60    &0.04     &-4.80    &-0.13    &-1.10    &-0.00     \\
-4.30    &-0.06    &-5.10    &0.06     &0.10     &-0.02    &-1.70    &0.08     &-1.80    &0.02     &10.10    &-0.03    &-5.10    &0.21     &0.50     &0.08     &-4.30    &0.04     &0.00   &-0.02     \\
-3.80    &0.01     &-4.60    &-0.39    &1.20     &-0.14    &-0.90    &-0.01    &-1.00    &0.24     &12.40    &-0.00    &-4.60    &0.06     &1.60     &0.06     &-3.80    &0.11     &1.10     &0.14      \\
-3.30    &0.08     &-4.20    &-0.19    &2.30     &0.06     &0.00   &0.31     &-0.10    &0.04     &14.60    &0.05     &-4.20    &0.18     &2.70     &-0.01    &-3.20    &0.05     &2.20     &-0.05     \\
-2.70    &0.08     &-3.70    &0.49     &3.50     &0.04     &0.90     &-0.02    &0.80     &0.08     &16.80    &-0.06    &-3.70    &0.16     &3.90     &-0.10    &-2.70    &-0.15    &3.20     &0.12      \\
-2.20    &-0.06    &-3.20    &-0.06    &4.60     &0.10     &1.80     &-0.21    &1.60     &-0.10    &...   &...   &-3.20    &-0.31    &5.00     &-0.02    &-2.20    &0.12     &4.30     &0.20      \\
-1.70    &-0.30    &-2.80    &0.04     &5.70     &0.26     &2.60     &0.05     &2.50     &0.01     &...   &...   &-2.80    &-0.17    &6.10     &0.09     &-1.60    &0.04     &5.40     &0.12      \\
-1.10    &0.22     &-2.30    &-0.49    &6.90     &0.24     &3.50     &0.07     &3.40     &-0.22    &...   &...   &-2.30    &-0.31    &7.20     &-0.11    &-1.10    &-0.12    &6.50     &0.22      \\
-0.60    &-0.33    &-1.90    &-0.25    &8.00     &-0.01    &4.40     &-0.02    &4.20     &-0.17    &...   &...   &-1.90    &-0.29    &8.30     &0.12     &-0.50    &-0.05    &7.50     &0.00      \\
-0.10    &0.31     &-1.40    &-0.08    &9.10     &0.05     &5.20     &0.47     &5.10     &0.01     &...   &...   &-1.40    &0.03     &9.50     &-0.05    &0.00   &-0.08    &8.60     &-0.08     \\
0.40     &-0.12    &-1.00    &0.03     &10.20    &0.02     &6.10     &0.31     &5.90     &0.46     &...   &...   &-0.90    &0.10     &10.60    &0.05     &0.50     &-0.06    &9.70     &0.03      \\
1.00     &0.22     &-0.50    &-0.09    &11.40    &0.05     &7.00     &0.34     &6.80     &0.09     &...   &...   &-0.50    &-0.31    &11.70    &-0.01    &1.10     &-0.11    &10.80    &0.07      \\
1.50     &-0.10    &-0.10    &-0.08    &12.50    &-0.02    &7.90     &-0.03    &7.70     &-0.02    &...   &...   &0.00   &-0.07    &12.80    &-0.06    &1.60     &-0.19    &11.80    &0.01      \\
2.00     &-0.26    &0.40     &0.48     &13.60    &0.04     &8.70     &0.05     &8.50     &0.03     &...   &...   &0.40     &0.45     &13.90    &0.01     &2.10     &0.14     &12.90    &0.06      \\
2.60     &0.07     &0.90     &-0.07    &14.70    &-0.05    &9.60     &-0.03    &9.40     &-0.02    &...   &...   &0.90     &-0.27    &15.10    &-0.02    &2.70     &0.09     &14.00    &0.00      \\
3.10     &0.34     &1.30     &-0.36    &15.90    &-0.07    &10.50    &0.20     &10.30    &-0.12    &...   &...   &1.30     &-0.06    &16.20    &-0.06    &3.20     &-0.04    &15.10    &0.08      \\
3.60     &0.29     &1.80     &-0.01    &17.00    &0.00     &11.30    &-0.09    &11.10    &-0.04    &...   &...   &1.80     &-0.07    &17.30    &0.00     &3.80     &0.16     &16.10    &0.06      \\
4.10     &0.67     &2.20     &0.44     &18.10    &-0.03    &12.20    &-0.01    &12.00    &-0.04    &...   &...   &2.30     &0.17     &18.40    &-0.05    &4.30     &0.51     &17.20    &0.08      \\
4.70     &0.60     &2.70     &0.28     &...   &...   &13.10    &-0.00    &12.90    &-0.20    &...   &...   &2.70     &-0.21    &...   &...   &4.80     &0.90     &...   &...    \\
5.20     &3.99     &3.10     &-0.19    &...   &...   &14.00    &-0.01    &13.70    &-0.16    &...   &...   &3.20     &0.02     &...   &...   &5.40     &1.59     &...   &...    \\
5.70     &7.73     &3.60     &0.13     &...   &...   &14.80    &0.18     &14.60    &0.07     &...   &...   &3.60     &-0.04    &...   &...   &5.90     &1.45     &...   &...    \\
6.30     &8.10     &4.00     &-0.12    &...   &...   &15.70    &-0.05    &15.50    &-0.09    &...   &...   &4.10     &0.03     &...   &...   &6.40     &1.67     &...   &...    \\
6.80     &4.84     &4.50     &0.19     &...   &...   &16.60    &-0.06    &16.30    &0.12     &...   &...   &4.60     &0.13     &...   &...   &7.00     &1.73     &...   &...    \\
7.30     &0.85     &5.00     &0.96     &...   &...   &17.40    &-0.05    &17.20    &0.07     &...   &...   &5.00     &0.82     &...   &...   &7.50     &0.65     &...   &...    \\
7.80     &0.18     &5.40     &2.32     &...   &...   &18.30    &0.06     &18.10    &-0.10    &...   &...   &5.50     &1.20     &...   &...   &8.00     &0.14     &...   &...    \\
8.40     &0.05     &5.90     &2.34     &...   &...   &...   &...   &18.90    &-0.01    &...   &...   &5.90     &1.29     &...   &...   &8.60     &0.06     &...   &...    \\
8.90     &0.49     &6.30     &3.11     &...   &...   &...   &...   &...   &...   &...   &...   &6.40     &1.57     &...   &...   &9.10     &0.02     &...   &...    \\
9.40     &-0.13    &6.80     &2.36     &...   &...   &...   &...   &...   &...   &...   &...   &6.80     &0.82     &...   &...   &9.70     &0.08     &...   &...    \\
10.00    &-0.07    &7.20     &0.81     &...   &...   &...   &...   &...   &...   &...   &...   &7.30     &0.37     &...   &...   &10.20    &0.05     &...   &...    \\
10.50    &0.07     &7.70     &-0.01    &...   &...   &...   &...   &...   &...   &...   &...   &7.80     &0.46     &...   &...   &10.70    &-0.16    &...   &...    \\
11.00    &0.28     &8.10     &0.41     &...   &...   &...   &...   &...   &...   &...   &...   &8.20     &-0.17    &...   &...   &11.30    &0.03     &...   &...    \\
11.50    &-0.11    &8.60     &-0.30    &...   &...   &...   &...   &...   &...   &...   &...   &8.70     &-0.14    &...   &...   &11.80    &0.02     &...   &...    \\
12.10    &-0.29    &9.00     &-0.32    &...   &...   &...   &...   &...   &...   &...   &...   &9.10     &-0.05    &...   &...   &12.30    &0.22     &...   &...    \\
12.60    &-0.04    &9.50     &0.15     &...   &...   &...   &...   &...   &...   &...   &...   &9.60     &-0.16    &...   &...   &12.90    &-0.01    &...   &...    \\
13.10    &0.06     &10.00    &0.53     &...   &...   &...   &...   &...   &...   &...   &...   &10.00    &0.18     &...   &...   &13.40    &-0.01    &...   &...    \\
13.70    &0.26     &10.40    &0.15     &...   &...   &...   &...   &...   &...   &...   &...   &10.50    &0.33     &...   &...   &14.00    &0.11     &...   &...    \\
14.20    &-0.05    &10.90    &-0.22    &...   &...   &...   &...   &...   &...   &...   &...   &11.00    &0.20     &...   &...   &14.50    &-0.07    &...   &...    \\
14.70    &-0.20    &11.30    &-0.22    &...   &...   &...   &...   &...   &...   &...   &...   &11.40    &-0.03    &...   &...   &15.00    &0.05     &...   &...    \\
15.20    &-0.25    &11.80    &0.01     &...   &...   &...   &...   &...   &...   &...   &...   &11.90    &-0.01    &...   &...   &15.60    &0.24     &...   &...    \\
15.80    &0.09     &12.20    &0.13     &...   &...   &...   &...   &...   &...   &...   &...   &12.30    &-0.12    &...   &...   &16.10    &0.18     &...   &...    \\
16.30    &0.14     &12.70    &0.04     &...   &...   &...   &...   &...   &...   &...   &...   &12.80    &0.25     &...   &...   &16.60    &-0.13    &...   &...    \\
16.80    &0.11     &13.10    &-0.14    &...   &...   &...   &...   &...   &...   &...   &...   &13.30    &0.23     &...   &...   &17.20    &-0.31    &...   &...    \\
17.40    &0.44     &13.60    &-0.36    &...   &...   &...   &...   &...   &...   &...   &...   &13.70    &0.01     &...   &...   &17.70    &-0.18    &...   &...    \\
17.90    &-0.08    &14.10    &-0.05    &...   &...   &...   &...   &...   &...   &...   &...   &14.20    &-0.16    &...   &...   &18.20    &-0.02    &...   &...    \\
18.40    &0.37     &14.50    &0.43     &...   &...   &...   &...   &...   &...   &...   &...   &14.60    &-0.14    &...   &...   &18.80    &0.02     &...   &...    \\
18.90    &-0.38    &15.00    &-0.11    &...   &...   &...   &...   &...   &...   &...   &...   &15.10    &0.31     &...   &...   &...   &...   &...   &...    \\
...   &...   &15.40    &0.23     &...   &...   &...   &...   &...   &...   &...   &...   &15.50    &0.52     &...   &...   &...   &...   &...   &...    \\
...   &...   &15.90    &0.30     &...   &...   &...   &...   &...   &...   &...   &...   &16.00    &-0.34    &...   &...   &...   &...   &...   &...    \\
...   &...   &16.30    &-0.31    &...   &...   &...   &...   &...   &...   &...   &...   &16.50    &0.00     &...   &...   &...   &...   &...   &...    \\
...   &...   &16.80    &0.26     &...   &...   &...   &...   &...   &...   &...   &...   &16.90    &0.42     &...   &...   &...   &...   &...   &...    \\
...   &...   &17.20    &-0.43    &...   &...   &...   &...   &...   &...   &...   &...   &17.40    &0.34     &...   &...   &...   &...   &...   &...    \\
...   &...   &17.70    &-0.05    &...   &...   &...   &...   &...   &...   &...   &...   &17.80    &0.35     &...   &...   &...   &...   &...   &...    \\
...   &...   &18.20    &-0.22    &...   &...   &...   &...   &...   &...   &...   &...   &18.30    &0.33     &...   &...   &...   &...   &...   &...    \\
...   &...   &18.60    &0.00     &...   &...   &...   &...   &...   &...   &...   &...   &18.80    &0.02     &...   &...   &...   &...   &...   &...    \\
\enddata
\end{deluxetable}

\begin{deluxetable}{lcccccc}
\tabletypesize{\scriptsize}
\tablecaption{Continuum flux in Jy and spectrally integrated line intensities and 2-$\sigma$ upper limits in Jy km s$^{-1}$ integrated over the disk area. \label{tbl:int}}
\tablewidth{0pt}
\tablehead{
\colhead{Species} & \colhead{DM Tau} & \colhead{AA Tau} &\colhead{LkCa 15} &\colhead{GM Aur} &\colhead{CQ Tau} &\colhead{MWC 480}
}
\startdata
218 GHz cont. &0.104[0.010]&0.073[0.007]&0.130[0.013]&0.177[0.018]&0.150[0.015]&0.255[0.026]\\
\smallskip
267 GHz cont. &0.143[0.014]&0.113[0.011]&0.214[0.021]&0.251[0.025]&0.258[0.026]&0.427[0.043]\\
        CO 2-1 &14.87[0.12]$^a$ & 8.20[0.16]$^a$ &13.94[0.15]$^a$ &19.41[0.11]$^a$ & 3.10[0.18]$^a$ &22.00[0.20]$^a$\\
   HCO$^+$ 3-2 & 5.34[0.12]$^a$ & 2.32[0.19]$^b$ & 5.19[0.17]$^a$ & 5.01[0.17]$^a$ & 0.75[0.24]$^a$ & 4.65[0.31]$^a$\\
   DCO$^+$ 3-2 & 0.71[0.04]$^b$ & $<$0.18$^a$ & 0.51[0.08]$^b$ & $<$0.15$^a$ & $<$0.23$^a$ & $<$0.31$^a$\\
N$_2$H$^+$ 3-2 & 0.97[0.07]$^c$ & (0.66[0.16])$^c$ & 0.71[0.14]$^c$ & 1.37[0.14]$^a$ &$<$0.35$^a$ & $<$0.75$^a$\\
   H$_2$CO 4-3 & (0.29[0.11])$^c$ & $<$0.52$^b$ & 1.12[0.22]$^b$ & 0.57[0.23]$^b$ &$<$0.42$^a$ &$<$0.78$^a$\\
   H$_2$CO 3-2 & 0.35[0.03]$^c$ & (0.16[0.08])$^b$ & 0.66[0.06]$^b$ & 0.56[0.08]$^b$ & $<$0.21$^a$ &$<$0.28$^a$\\
       HCN 3-2 & 2.92[0.11]$^b$ & 0.59[0.13]$^c$ & 5.52[0.21]$^a$ & 1.27[0.12]$^c$ & $<$0.50$^a$ & 2.34[0.19]$^b$\\
       DCN 3-2 &$<$0.10$^a$& $<$0.18$^a$ &0.41[0.05]$^b$ & $<$0.09$^a$ & $<$0.17$^a$ & $<$0.22$^a$\\
CN 2$_3$-1$_2$ & 4.74[0.08]$^a$ & 1.76[0.09]$^b$ & 6.80[0.11]$^a$ & 0.96[0.08]$^a$ & 0.22[0.11]$^b$ & 3.29[0.14]$^a$\\
CN 2$_2$-1$_1$ & 0.57[0.07]$^a$ & [0.08]$^b$ & 1.16[0.12]$^a$ & $<$0.19$^a$ & $<$0.22$^b$ & 0.61[0.15]$^a$\\
\enddata
\\The integrated intensities are extracted from $^a$ the complete CO disk, $^b$ 3/4 of the CO disk and $^c$ 1/2 of the CO disk. The 1-$\sigma$ uncertainties in the integrated intensities are in square parentheses. Intensities in parentheses are tentative detections.
\end{deluxetable}

\end{document}